\documentclass[aps,pre,amsmath,twocolumn,superscriptaddress,showpacs]{revtex4-1}
\usepackage{calc}
\usepackage{amsmath}
\usepackage{amsfonts}
\usepackage{amssymb}
\usepackage{graphicx}
\usepackage{hyperref}
\usepackage{soul}
\usepackage{mathtools}
\usepackage{verbatim}
\usepackage{epstopdf}
\usepackage{subfigure}
\usepackage{latexsym}
\usepackage{dcolumn}
\usepackage{epsf}
\usepackage{float}
\usepackage[table]{xcolor}
\usepackage{multirow}

\usepackage{color} 

\usepackage{graphicx}
\usepackage{graphicx,epstopdf}
\usepackage{xcolor}
\usepackage{dcolumn}
\usepackage{bm}
\usepackage{mathrsfs} 
\newtheorem{theorem}{Theorem}

\begin{document}
\title{Swarmalators under competitive time-varying phase interactions}
	
\author{Gourab K. Sar}
\affiliation{Physics and Applied Mathematics Unit, Indian Statistical Institute, 203 B. T. Road, Kolkata 700108, India}
\author{Sayantan Nag Chowdhury}
\affiliation{Physics and Applied Mathematics Unit, Indian Statistical Institute, 203 B. T. Road, Kolkata 700108, India}
\author{Matja{\v z} Perc}
\affiliation{Faculty of Natural Sciences and Mathematics, University of Maribor, Koro{\v s}ka cesta 160, 2000 Maribor, Slovenia}
\affiliation{Department of Medical Research, China Medical University Hospital, China Medical University, Taichung, Taiwan}
\affiliation{Complexity Science Hub Vienna, Josefst{\"a}dterstra{\ss}e 39, 1080 Vienna, Austria}
\affiliation{Alma Mater Europaea, Slovenska ulica 17, 2000 Maribor, Slovenia}
\author{Dibakar Ghosh}
\email{dibakar@isical.ac.in}
\affiliation{Physics and Applied Mathematics Unit, Indian Statistical Institute, 203 B. T. Road, Kolkata 700108, India}	


	

	\begin{abstract}
	
	
	Swarmalators are entities with the simultaneous presence of swarming and synchronization that reveal emergent collective behavior due to the fascinating bidirectional interplay between phase and spatial dynamics. Although different coupling topologies have already been considered, here we introduce time-varying competitive phase interaction among swarmalators where the underlying connectivity for attractive and repulsive coupling varies depending on the vision (sensing) radius. Apart from investigating some fundamental properties like conservation of center of position and collision avoidance, we also scrutinize the cases of extreme limits of vision radius. The concurrence of attractive-repulsive competitive phase coupling allows the exploration of diverse asymptotic states, like static $\pi$, and mixed phase wave states, and we explore the feasible routes of those states through a detailed numerical analysis. In sole presence of attractive local coupling, we reveal the occurrence of static cluster synchronization where the number of clusters depends crucially on the initial distribution of positions and phases of each
	swarmalator. In addition, we analytically calculate the sufficient condition for the emergence of the static synchronization state. We further report the appearance of the static ring phase wave state and evaluate its radius theoretically. Finally, we validate our findings using Stuart-Landau oscillators to describe the phase dynamics of swarmalators subject to attractive local coupling.

\end{abstract}

\maketitle
\section{Introduction}
One of the most common attributes among different organisms in nature is to dwell in groups or move in consensus and mimic the activities of their local neighbors, the reason of which can be traced back to the survival instinct of those organisms. The examples of which can be found in systems as small as bacterial aggregation \cite{levy2008stochastic, chavy2016local} to macro-organisms such as flock of birds, herd of sheep, and school of fish \cite{bialek2012statistical, garcimartin2015flow, barbaro2009discrete, couzin2007collective, sumpter2010collective, herbert2016understanding}. In all these systems, the individuals organize their positions in space to aggregate together or move in unison. This phenomenon, commonly known as swarming \cite{fetecau2011swarm, topaz2004swarming, mogilner1999non, reynolds1987flocks}, is widespread in coordinated movement of a group of animals. Swarming usually means self-organization of entities in space without considering the effect of the internal state. Another such collective behavior is synchronization \cite{winfree1967biological,uzuntarla2019synchronization, kuramoto1975international,chowdhury2019convergence,anwar2022stability, pikovsky2001universal,rakshit2021relay,uzuntarla2019firing, mirollo1990synchronization,chowdhury2019synchronization1}, which is more ubiquitous in nature and technology, where the units adjust their internal states to self-organize in time. Flashing of fireflies \cite{buck1988synchronous}, chorusing frogs \cite{aihara2008mathematical}, firing neurons \cite{rakshit2018synchronization, montbrio2015macroscopic}, phase-locking in Josephson junction \cite{wiesenfeld1996synchronization, vlasov2013synchronization} are some of the well-known instances where synchronization occur. Here, only the oscillator's internal phase dynamics receives the central focus without shedding much light on the spatial motion. Examples are also found in nature where oscillator's spatial and phase dynamics affect each other \cite{riedel2005self, yan2012linking, nguyen2014emergent}. Tree frogs, crickets, katydids synchronize their calling rhythms with nearby individuals, and their movements are believed to be influenced by relative phases of their calling \cite{walker1969acoustic, greenfield1994synchronous, greenfield1993katydid}. The study of ferromagnetic colloids, sperms, land-based robots, aerial drones, and other active entities involves both dynamics \cite{snezhko2011magnetic, yang2008cooperation, barcis2019robots, barcis2020sandsbots}.

\par Particles whose spatial and phase dynamics affect each other are commonly known as {\it swarmalators} \cite{o2017oscillators} in the essence that they swarm in space to self-organize their positions and simultaneously oscillate to adjust their internal state. Earlier, in the study of {\it mobile agents} or {\it moving oscillators}, the influence of agents' motion on their phase dynamics are considered, but their spatial dynamics are not affected by phase dynamics \cite{vicsek1995novel, stilwell2006sufficient, frasca2008synchronization, chowdhury2019synchronization, majhi2019emergence}. The first step towards the study of swarmalators was taken by Tanaka and others while describing the diverse phenomena of chemotactic oscillators \cite{tanaka2007general, iwasa2010hierarchical}. Recently, O'Keeffe et al. \cite{o2017oscillators} modeled the swarmalators by incorporating appropriate coupling functions where the influence of spatial and phase dynamics on each other was aptly illustrated. Swarmalator model for global interaction among the units is governed by the pair of equations \cite{o2018ring},

\begin{equation}
	\label{eq.1}
	\dot{\textbf{x}}_{i} = \textbf{v}_{i}+\frac{1}{N-1} \sum_{\substack{j = 1\\j \neq i}}^{N}\bigg[\text{I}_{\text{att}}(\textbf{x}_{ij}) \text{F}_{\text{att}}(\theta_{ij}) - \text{I}_{\text{rep}}(\textbf{x}_{ij}) \text{F}_{\text{rep}}(\theta_{ij})\bigg],
\end{equation}

\begin{equation}
	\label{eq.2}
	\dot{\theta}_{i} = \omega_{i}+\frac{\epsilon}{N-1}\sum_{\substack{j = 1\\j \neq i}}^{N} \text{H}(\theta_{ij})\text{G}(\textbf{x}_{ij}),
\end{equation}
for $i = 1,2,\ldots,N$, where $N$ is the total number of swarmalators, $\textbf{x}_i = (x_i, y_i) \in \mathbb{R}^2$ is the position vector of the $i$-th swarmalator, $\theta_i \in \mathbb{S}^1$ is its internal phase, while $\textbf{v}_i$ and $\omega_i$ denote its self-propulsion velocity and natural frequency, respectively with $\textbf{x}_{ij} \equiv \textbf{x}_j - \textbf{x}_i$ and $\theta_{ij} \equiv \theta_j - \theta_i$. The spatial attraction and repulsion are governed by the functions $\text{I}_\text{att}$ and $\text{I}_\text{rep}$. $\text{F}_{att}$ and $\text{F}_\text{rep}$ represent the influence of phase similarity on spatial attraction and repulsion, respectively. In Eq.\ \eqref{eq.2}, phase interaction between the swarmalators is controlled by the function $\text{H}$ and the influence of spatial proximity on phase dynamics is given by $\text{G}$. Swarmalators' phases are coupled with strength $\epsilon \in \mathbb{R}$. When $\epsilon$ is positive (attractive coupling), swarmalators try to minimize their phase differences while a negative value (repulsive coupling) of $\epsilon$ increases the incoherence of their phases. In Ref.~\cite{o2017oscillators}, three stationary (one each for attractive, repulsive, and absence of phase coupling) and two non-stationary states (both for repulsive phase coupling) of possible long-term aggregation are found. More new states are reported when the model of swarmalators is extended by addition of periodic forcing \cite{lizarraga2020synchronization}, noise \cite{hong2018active}, and finite cut-off interaction distance \cite{lee2021collective, jimenez2020oscillatory}. Finite size effects \cite{o2018ring} of the population of swarmalators and the well-posedness of solution in the mean-field limit \cite{ha2021mean, ha2019emergent} are also studied. 
The collective behavior of the swarmalators positioned on the one-dimensional ring is studied analytically in Ref.\ \cite{o2021collective}. The sign of phase coupling determines the coherent or incoherent nature of the states. 

\par Mixed influence of positive and negative couplings can be found in interactions in neuronal networked systems \cite{hopfield1982neural} and in the calling behavior of Japanese frogs \cite{aihara2008mathematical}. Coupling disorder with random interaction of swarmalators' phases is introduced in Ref.~\cite{hong2021coupling}, where chimera-like states are observed. The nature of coupling in most real world systems is often complex, which motivates us to study the behavior of swarmalators under the mixed coupling strategy. More importantly, earlier, most of the existing studies on coexisting attractive-repulsive interaction \cite{hong2011kuramoto,majhi2020perspective,hong2011conformists,chowdhury2021antiphase,yuan2018periodic,chowdhury2020effect} were performed on static network formalism. Such signed networks can display fascinating macroscopic dynamics \cite{hong2011kuramoto,iatsenko2014glassy,iatsenko2013stationary}, including the $\pi$ state, the traveling wave state, and the mixed state. 
Recently, the impact of such competitive interactions through the concurrence of positive-negative coupling has been investigated on the time-evolving networks of mobile agents \cite{chowdhury2019synchronization,chowdhury2020distance}, leading to diverse peculiar dynamical states, including extreme events \cite{chowdhury2021extreme,chowdhury2021extrememap}. However, these studies consider only the unidirectional influence of spatial dynamics towards the oscillator's amplitude and phase dynamics. This present article focuses on the bidirectional interplay between swarmalators' phase values and spatial positions. To the best of our knowledge, in spite of the colossal importance of time-varying interaction \cite{holme2012temporal,nag2020cooperation,majhi2017amplitude,dixit2021dynamic,ghosh2022synchronized,dixit2021emergent,majhi2017synchronization} from diverse aspects, the study of swarmalators is less explored, particularly during the simultaneous presence of
attractive-repulsive temporal interaction in the phase dynamics.

\par We are curious to investigate how competitive phase interaction induces long-term states of position and phase aggregation of the system. We design a particular coupling scheme to serve this purpose. A communication circle with a fixed and uniform radius is associated with each swarmalator so that the swarmalators within this vision range interact bidirectionally with positive coupling strength. Outside this attractive vision range, each swarmalator goes through repulsive phase coupling. Under this setup, suitable choices of parameters lead to various emergent states due to the interplay between attractive and repulsive phase couplings among the swarmalators with global spatial attraction and repulsion. We emphasize the role of attractive vision radius in realizing the collective dynamics and examine the possible routes for achieving the static sync state. The extreme limit of this attractive vision radius transforms the phase interaction into a global attractive or repulsive one and leads to different emergent patterns like static sync, static async, active phase wave, and splintered phase wave. We elaborately discuss the main features of these patterns depending on the phase-dependent spatial dynamics and position-dependent phase dynamics. The absence of repulsive coupling strength ensures the manifestation of static cluster synchronization. We are able to derive a sufficient condition for the transition from incoherence to static sync state. We also identify a novel state, viz.\ static inscribed cluster with local attractive coupling. When $F_{rep}$ depends on the phase dynamics explicitly, we trace out the static ring phase wave state under the influence of global repulsive coupling. We analytically derive the radius of this stationary state and validate the theoretical findings through numerical simulation. Carrying forward our interest in local phase interaction, we model swarmalators where the Stuart-Landau limit-cycle oscillators govern the phase dynamics.

\par The subsequent sections of this article are arranged as follows. In Sec.\ \ref{sec:level2}, we introduce a two-dimensional swarmalator model endowed with global co-existing spatial attraction-repulsion and competitive phase interaction. Section\ \ref{sec:level3} contains theoretical analysis where we ensure the non-existence of finite time collision and the existence of a minimal inter-particle distance between the swarmalators. We discuss in detail the main findings of our investigation in Sec.\ \ref{sec:level4}. The emergence of different collective behaviors and their characterization with order parameters are studied by considering several cases. Section\ \ref{sec:level5} includes the study of our model when the Stuart-Landau oscillator \cite{kuramoto2003chemical} is used for the phase dynamics with local attractive phase coupling. Finally, we summarize our work in Sec.\ \ref{sec:level6} and discuss the possible scope for future research. We include Appendix A \eqref{sec:level9} dealing with the swarmalators moving in three-dimensional space to inspect the impact of higher dimensional spatial dynamics on our proposed model.


\section{\label{sec:level2}Swarmalator model with competitive phase interaction}
\par
\begin{figure*}[hpt]
	\centerline{
		\includegraphics[scale = 0.70]{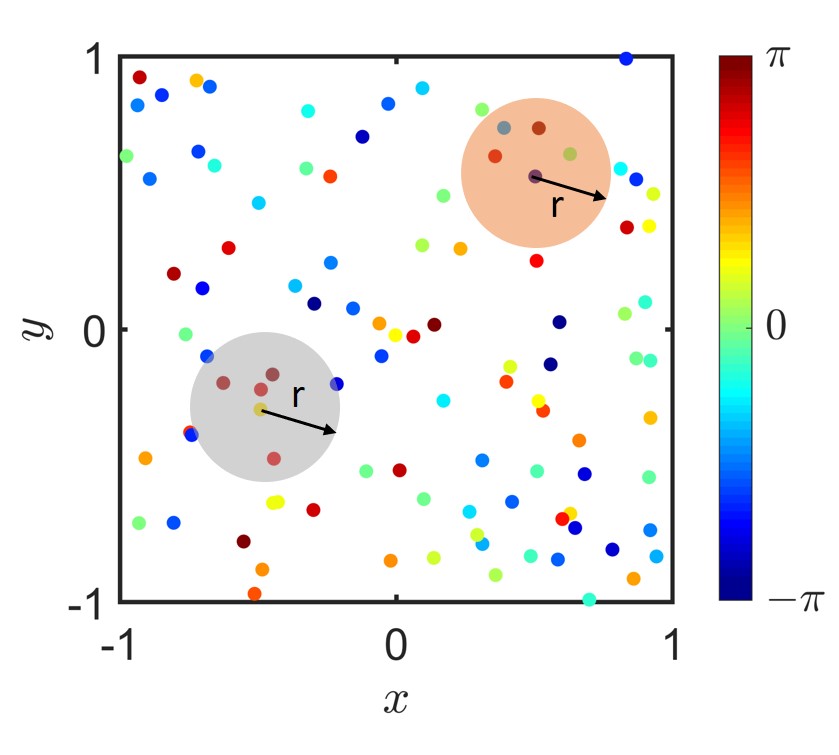}}
	\caption{{\bf Schematic diagram of $N = 100$ swarmalators with circular attractive range of vision}: The swarmalators positioned randomly in the two-dimensional plane inside the bounded region $[-1,1] \times [-1,1]$ are represented by the small dots which are colored according to their phases, drawn from $[-\pi, \pi]$ uniformly at random. Two circular regions are the respective attractive vision ranges for two distinct particles, where the central swarmalators are attractively phase coupled with all other swarmalators inside and is repulsively phase coupled with all those who are outside the regions.}
	\label{Fig.1}
\end{figure*}

 Every swarmalator moves in space with an attractive vision radius $r$, and the closed circular region it covers being at the center, can be considered as its attractive range of interaction. In a sense, they can feel the presence of nearby swarmalators and couple their phases attractively with positive coupling strength. Outside the attractive vision range the positive connectivity gets lost, and they are coupled repulsively (see Fig.\ \ref{Fig.1}). The attractive-repulsive phase coupling and the all-to-all spatial attraction and repulsion make the swarmalator model more challenging as the competitive phase interaction increases the possibility of finding new states.  Instead of choosing this circular vision shape of a uniform radius, one can select other polygons and heterogeneous communication radii. However, the results shown in this manuscript remain the same qualitatively for such choices. We propose the following model,

\begin{equation}
	\label{eq.3}
	\dot{\textbf{x}}_{i} = \textbf{v}_{i}+\frac{1}{N-1} \sum_{\substack{j = 1\\j \neq i}}^{N}\left[ \frac{\textbf{x}_{ij}}{|\textbf{x}_{ij}|^ \alpha} (1+J\cos(\theta_{ij}))  -    \frac{\textbf{x}_{ij}}{{|\textbf{x}_{ij}|}^{\beta}} \right],
\end{equation}

\begin{equation}	
	\label{eq.4}
	\dot{\theta}_{i} = \omega_{i}+\frac{\epsilon_a}{N_{i}}\sum_{\substack{j = 1\\j \neq i}}^{N} \text{A}_{ij}\frac{\sin(\theta_{ij})}{|\textbf{x}_{ij}|^\gamma} +  \frac{\epsilon_r}{N-1-N_{i}}\sum_{\substack{j = 1\\j \neq i}}^{N} \text{B}_{ij}\frac{\sin(\theta_{ij})}{|\textbf{x}_{ij}|^ \gamma}.
\end{equation}

Here, we consider the set $\Lambda_{i}(r) = \{j \in \{1,2,...,N\}\; \mbox{such that} \; |\textbf{x}_{ij}| \le r, j \ne i\}$ and $N_{i}$ is the number of elements in $\Lambda_{i}(r)$. $N_{i} \ne 0$ means there is at least one swarmalator inside the attractive vision range of the $i$-th swarmalator except itself and $N_{i} \ne N-1$ indicates the presence of at least one swarmalator outside its attractive vision range. $\text{A}=[\text{A}_{ij}]_{N \times N}$ and $\text{B}=[\text{B}_{ij}]_{N \times N}$ are the adjacency matrices for attractive and repulsive phase couplings, respectively, and are defined as,
\begin{equation*}
	\text{A}_{ij} =
	\begin{cases}
		1 & \text{if $j \in \Lambda_{i}(r)$}\\
		0 & \text{otherwise}
	\end{cases}
	;\hspace{5pt} \text{B}_{ij} =
	\begin{cases}
		1 & \text{if $j \not\in \Lambda_{i}(r) \cup \{i\}$}\\
		0 & \text{otherwise}
	\end{cases}
\end{equation*}

Equation \eqref{eq.4} remains well-defined, if $N_{i}$ $\neq$ $0$ and $(N-1)$. When $N_{i}$ attains these two values $0$ and $(N-1)$, then the matrices $\text{A}$ and $\text{B}$ become null matrices, respectively. We will study two extreme cases for $N_{i}$ $=$ $0$ and $N_{i} = (N-1)$ later in the subsection~\ref{subsec:level1} in terms of attractive vision radius $r$. We choose power law attraction and repulsion for $\text{I}_{\text{att}}$ and $\text{I}_{\text{rep}}$ with positive exponents $\alpha$ and $\beta$, respectively. Note that, Eq.\ \eqref{eq.3} can be written as
\begin{equation}
	\dot{\textbf{x}}_{i} = \textbf{v}_{i}+\frac{1}{N-1} \sum_{\substack{j = 1\\j \neq i}}^{N} \text{F}(|\textbf{x}_{ji}|) \frac{\textbf{x}_{ji}}{|\textbf{x}_{ji}|},
	\label{eq.5}
\end{equation}

where $\text{F}(|\textbf{x}_{ji}|) = \frac{1}{|\textbf{x}_{ji}|^{\beta-1}} - \frac{1+J\cos(\theta_{ji})}{|\textbf{x}_{ji}|^{\alpha-1}}$ is the net force exerted on the $i$-th swarmalator by the $j$-th one, along the direction $\textbf{x}_{ji}$. When this force is negative, swarmalators attract each other following Eq.\ \eqref{eq.5} and similarly positive force means repulsion between them. As a result, this force must be positive (or repulsive) for nearby swarmalators so that they do not collide, whereas it must be negative (or attractive) for swarmalators far away from each other so that they do not disperse indefinitely \cite{fetecau2011swarm}. To ensure this, we need to choose $\alpha$ and $\beta$ such that $1\le \alpha < \beta$ holds. In our work, without loss of any generality, we consider the influence of phase only on spatial attraction (i.e., spatial repulsion between swarmalators is independent of their phases) by our choice of functions, $\text{F}_{\text{att}}(\theta) = 1+J\cos(\theta)$ and $\text{F}_{\text{rep}} = 1$.  However, we also study the behavior of swarmalators when phase similarity affects spatial repulsion for a particular case in Sec.\ \eqref{sec:level7}. The parameter $J$ measures how phase similarity influences spatial attraction. We choose $0 \leq J <1$, so that the attraction function is always positive. Positive value of $J$ indicates that swarmalators which are in nearby phases, attract themselves spatially and stay close to each other.  If $J$ is negative, swarmalators are spatially attracted to all those in opposite phases. In Sec.\ \eqref{sec:level8}, we study our model by considering negative values of $J$ under the presence of local attractive coupling only. The phase interaction function $\text{H}$ is taken as the sine function inspired by the Kuramoto model \cite{kuramoto1975international}. To capture the spatial influence of the swarmalators on their phase dynamics, we choose $\text{G}(\textbf{x}) = \frac{1}{|\textbf{x}|^ \gamma} (\gamma > 0)$. Results in this paper are presented with a specific choice of parameters' values, viz.\ $\alpha = 1$, $\beta = 2$, and $\gamma = 1$, unless otherwise mentioned.  These choices of $\alpha$, $\beta$, and $\gamma$ make sure that the solutions are bounded and prevent inter-particle collision. 
For simplicity, we choose identical swarmalators so that $\textbf{v}_i = \textbf{v}$ and $\omega_{i} = \omega$, and by proper choice of reference frame we set both $|\textbf{v}|$ and $\omega$ to zero.  In Fig.\ \ref{Fig.1}, a schematic diagram of the initial positions of the swarmalators in the two-dimensional plane is shown, where they are colored according to their initial phases. Initially, the swarmalators are positioned randomly inside the region $[-1,1] \times [-1,1]$ and their phases are selected randomly from $[-\pi,\pi]$.

\par It should be noted that the effect of spatial position on the internal phase has two folds. The swarmalator's position in space controls the phase coupling strategy as well as the effective interaction strength. The positions of the swarmalators in the two-dimensional plane determine whether their phases are coupled attractively or repulsively, depending on the attractive vision radius $r$. Also, the strength of phase interaction is multiplied by $\frac{1}{\textbf{x}_{ij}}$ which indicates that the swarmalators staying nearby in space are coupled with more strength than swarmalators who stay some distance apart. The parameters $\epsilon_a$ ($>0$) and $\epsilon_r$ ($<0$) control the strength of attractive and repulsive phase interactions, respectively. To avoid monotony, from hereon by vision radius and vision range, we will mean attractive vision radius and attractive vision range, respectively. For the sake of simplicity, we also include Table \ref{tab:table1} mentioning all the observed states and their corresponding abbreviations used in this article.

\begin{table}
	\caption{\label{tab:table1} Various emergent collective states and their abbreviations:}
	\centerline{
		\begin{tabular}{ccccc}
			
			Emergent collective states & Abbreviations\\ \hline
			Static sync& SS \\
			Static async& SA \\
			Static phase wave& STPW \\
			Splintered phase wave& SPPW \\
			Active phase wave& APW \\
			Static cluster sync& SCS \\
			Static $\pi$ & SPI \\
			Attractive mixed phase wave& AMPW \\
			Repulsive mixed phase wave& RMPW \\
			Static inscribed cluster & SIC\\
			Static ring phase wave & SRPW
	\end{tabular}}
\end{table}



\section{\label{sec:level3}Theoretical Analysis}
The proposed model defined by Eqs.\ \eqref{eq.3} and \eqref{eq.4} captures the spatial and phase dynamics of swarmalators where they are interconnected. This interplay between swarmalators' position and phase gives rise to different complex dynamical patterns which we will discuss in the following sections. Before moving into illustrating the asymptotic states of the system, first we discuss some basic properties of the proposed model. Let $\Gamma(\textbf{x}_{ij},\theta_{ij}) = \frac{\textbf{x}_{ij}}{|\textbf{x}_{ij}|^ \alpha} (1+J\cos(\theta_{ij}))  -    \frac{\textbf{x}_{ij}}{{|\textbf{x}_{ij}|}^{\beta}}$. So, Eq.\ (\ref{eq.3}) with $\textbf{v}_{i}=\bf{0}$ can be rewritten as,
\begin{equation}
	\label{eq.6}
	\dot{\textbf{x}}_{i} = \frac{1}{N-1} \sum_{\substack{j = 1\\j \neq i}}^{N} \Gamma(\textbf{x}_{ij},\theta_{ij}).
\end{equation}
It is easy to notice that, $\Gamma(\textbf{x}_{ij},\theta_{ij}) = - \Gamma(\textbf{x}_{ji},\theta_{ji})$, i.e., $\Gamma$ is skew-symmetric under the exchange of indices $i$ and $j$. If we take the total sum by considering all the particles, then it will be identically zero. As a result 
\begin{equation}
	\label{eq.7}
	\sum_{\substack{i = 1}}^{N}\dot{\textbf{x}}_{i} = \frac{1}{N-1} \sum_{\substack{i = 1}}^{N}\sum_{\substack{j = 1\\j \neq i}}^{N} \Gamma(\textbf{x}_{ij},\theta_{ij}) = 0.
\end{equation}
So, the center of position of the system $\frac{1}{N}\sum_{\substack{i = 1}}^{N}{\textbf{x}}_{i}$ is always conserved. Hence, the arithmetic mean of the initial distribution of the spatial position of the system at $t=0$ from the box $[-1,1] \times [-1,1]$ remains invariant for future iterations. However, the conservation of mean phase $\frac{1}{N}\sum_{\substack{i = 1}}^{N}\theta_{i}$ can not be assured, as the phase interaction in Eq.\ (\ref{eq.4}) is split into two different components, viz.\ attractive and repulsive parts. These two separate portions do not always combine resulting into the vanishing of the total sum $\sum_{\substack{i = 1}}^{N}\dot{\theta}_{i}$.

\par One of the main difficulties in studying the system theoretically is the presence of singular terms like $|\textbf{x}_{ij}|^ \eta$ in the denominators of the coupling functions. The question that arises is whether the system is well-defined for the case of a collision at any finite time between two or more swarmalators or not.	
To encounter this, we chose coupling functions so that the finite time collision avoidance between the swarmalators can be assured. To avoid notation complexity, we set $\mathcal{N} = \{1,2,...,N\}$.


\begin{theorem}\label{Theo:1}
	Suppose $1 \le \alpha < \beta$ and the initial data is chosen for which there is no collision among the swarmalators, i.e., for all $i,j \in \mathcal{N}$ and $i \ne j$,
	\[ \min_{1\le i,j \le N} |\mathbf{x}_{i}(0)-\mathbf{x}_{j}(0)| > 0. \]
	Then, in finite time, the swarmalators will never collide, i.e., there exists a global solution of Eqs.\ \eqref{eq.3} and \eqref{eq.4} with
	\[\mathbf{x}_{i}(t) \ne \mathbf{x}_{j}(t),\]
	for all $i \ne j$, $t \in (0, \infty)$. Moreover, there exists a positive lower bound for the minimal inter-particle distance, $\delta$ such that,
	\[\inf_{0\le t< \infty} \min_{i,j} |\textbf{x}_{i}(t)-\textbf{x}_{j}(t)| \ge \delta.\]
\end{theorem}

\paragraph*{ Sketch of the proof:}
This theorem mainly focuses on how the swarmalators move in space without colliding with each other. To prove these results, we have adopted the method of Ref.~\cite{ha2019emergent} where a general swarmalator model with global space and phase coupling was considered. Collision avoidance and minimal inter-particle distance are spatial properties of the swarmalators, irrespective of their phase. Eq.\ (\ref{eq.3}) of our model corresponds to the spatial dynamics of (1.1) in Ref.~\cite{ha2019emergent} with the choices $\omega_i = 0, \forall i \in \mathcal{N}$, $\Gamma_a(\theta) = 1+J\cos(\theta)$, and $\Gamma_r(\theta) = 1$. The main difference of our model given by Eqs.\ \eqref{eq.3}-\eqref{eq.4} and the one in Ref.~\cite{ha2019emergent} is in the phase dynamics of the swarmalators. Since the phase equation does not play any role in proving these results, we can proceed in the same way to prove this theorem.

The last theorem eliminates the possibility of collision among particles in finite time. So, the solution of this system is always well-posed. Now, we can proceed to study the collective states of their position aggregation and phase synchronization. To investigate the phase synchrony of the swarmalators, we use the complex order parameter of Kuramoto model defined by

\begin{equation}
	R e^{l \bar{\Theta}} = \frac{1}{N} \sum_{\substack{j = 1}}^{N} e^{l \theta_j}, (l = \sqrt{-1})
	\label{eq.8}
\end{equation}

where $0 \le R \le 1$ measures the coherence of swarmalators' phases and $\bar{\Theta}$ is their mean phase. Here $R \ll 1$ represents there is no convergence into a unique phase among the swarmalators. On the other hand, when swarmalators' phases are fully synchronized, then we have complete phase coherence with $R = 1$.

\par Swarmalators organize themselves in space and adjust their phases with others. The correlation between swarmalator's phase $\theta$ and spatial angle $\phi = \tan ^{-1}(y/x)$ is measured by another order parameter defined by

\begin{equation}
	S_{\pm} e^{l \Psi_{\pm}} = \frac{1}{N} \sum_{\substack{j = 1}}^{N} e^{l(\phi_j \pm \theta_j)}.
	\label{eq.9}
\end{equation}
When the phase and spatial angle of each swarmalator is perfectly correlated, we have $\phi_i = \pm \theta_i + C$ (for some constant $C$ which depends on the initial conditions). In this case, we have $S_\pm = 1$, and this value of $S_\pm$ decreases when the correlation is reduced. The maximum of $S_\pm$ is defined by $S = \max(S_+,S_-)$  and can be chosen to measure this correlation.

\par For numerical simulation in our study, we use FORTRAN 90 compiler and the integration is done using Runge Kutta method of order 4 with step size 0.01. Swarmalators are initially distributed inside the bounded region $[-1,1] \times [-1,1]$ and their phases are drawn from $[-\pi, \pi]$ randomly.

\begin{figure*}[hpt]
	\includegraphics[scale = 0.36]{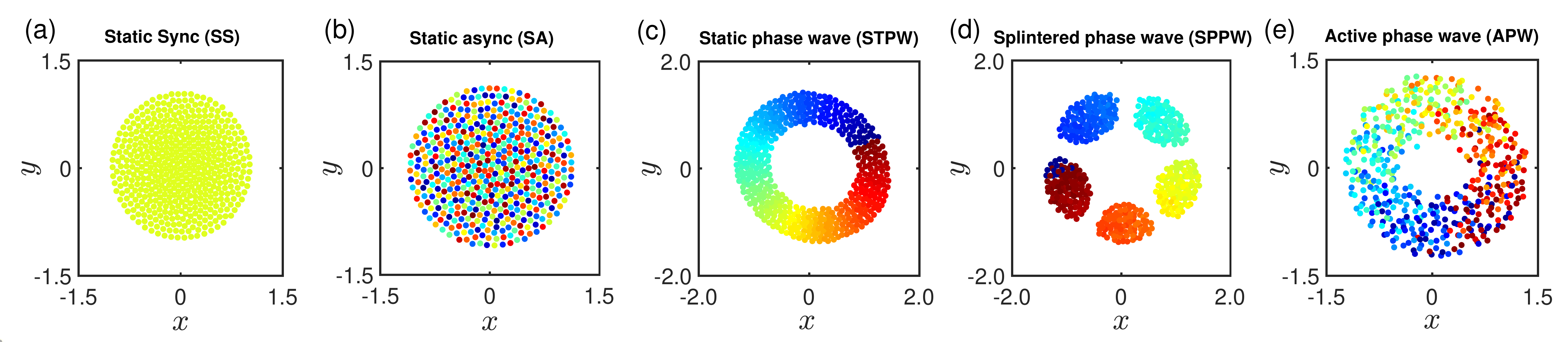}					
	\caption{{\bf Scatter plots of swarmalators in the $(x,y)$ plane for different static and active states}: Here, the term `active' reflects movement in space and continuous change in the phases of the swarmalators in the asymptotic states. In contrast, the term `static' demonstrates stationarity in spatial as well as in the phase dynamics.  The numerical simulations are done with $N = 500$ swarmalators over 30000 iterations with integration step size 0.01. The snapshots shown here are taken at the last $(300)$ time unit. (a) Static sync (SS) for $(J, r, \epsilon_a, \epsilon_r) = (0.1,5.0,0.5,-0.5)$, (b) static async (SA) for $(J, r, \epsilon_a, \epsilon_r) = (0.1, 10^{-5}, 0.5, -0.5)$, (c) static phase wave (STPW) for $(J, r, \epsilon_a, \epsilon_r) = (0.9, 10^{-5}, 0.5, 0.0)$, (d) splintered phase wave (SPPW) for $(J, r, \epsilon_a, \epsilon_r) = (0.9, 10^{-5}, 0.5, -0.1)$, and (e) active phase wave (APW) for $(J, r, \epsilon_a, \epsilon_r) = (0.75, 10^{-5}, 0.5, -0.5)$. We classify all these states depending on their respective phase values as well as spatial positions. Please see the main text for detailed descriptions of these states. }
	\label{Fig.2}
\end{figure*}

\section{\label{sec:level4}Results}

In this section, we consider different coupling topologies. The behaviors of the swarmalators when vision radius is very large and very small are studied in Sec.~\ref{subsec:level1}. Local attractive coupling is considered in Sec.~\ref{subsec:level2} for phase interaction among swarmalators. Finally, in Sec.~\ref{subsec:level3}, asymptotic states of the swarmalators under attractive-repulsive phase coupling are discussed.

\subsection{\label{subsec:level1}Extreme limits of vision radius ($r$)}

First, we want to study the two cases when the vision range of the swarmalators is either very large or very small. When the swarmalators move in space with an infinite vision radius, they will sense the presence of every other swarmalators within its vision range. In that case, $\Lambda_i (r) = \{1,\ldots,i-1,i+1,\ldots,N\}$ and as a result the repulsive matrix $\text{B}$ becomes null. This means swarmalators' phases are attractively coupled only with the coupling strength $\epsilon_a >0$. So, the phase dynamics given by Eq.\ \eqref{eq.4} with $\omega_i = 0$ effectively becomes,
\begin{equation}	
	\dot{\theta}_{i} =\frac{\epsilon_a}{N-1}\sum_{\substack{j = 1\\j \neq i}}^{N}\frac{\sin(\theta_{ij})}{|\textbf{x}_{ij}|^\gamma} \hspace{0.2cm}(\epsilon_a>0).
	\label{eq.10}
\end{equation}
The positive phase coupling strength along with the global interaction (note that, the attractive coupling matrix $\text{A}$ becomes $\text A_{ij} = 1$ ($i \ne j$), $\text A_{ii} = 0$ for $i,j \in \mathcal{N}$) makes the swarmalators minimize their phase difference. Phase coherence among the swarmalators is found, where they lie
inside a circular disc in two-dimensional plane. The formation of this disc structure is influenced by the force function $\text{F}$ (cf.\ Eq.\ \ref{eq.5}), which leads to uniform density of particles inside a disc in the absence of $J$. Due to the complete coherence in swarmalators' phases, the phase influence on spatial position $1+J\cos(\theta_{ij})$ becomes $1+J$, a constant, and the disc-like structure is sustained with the radius being scaled by $\frac{1}{1+J}$. The order
parameter R gives the value 1 here, justifying the swarmalators' totally phase
synchronized state. In sufficiently long time, swarmalators stop moving in space, and remain spatially static. Besides, their phases become static as well. This state is called static synchrony (SS) (See Fig.\ \ref{Fig.2}(a)).

\begin{figure*}[hpt]
	\centerline{
		\includegraphics[scale = 0.85]{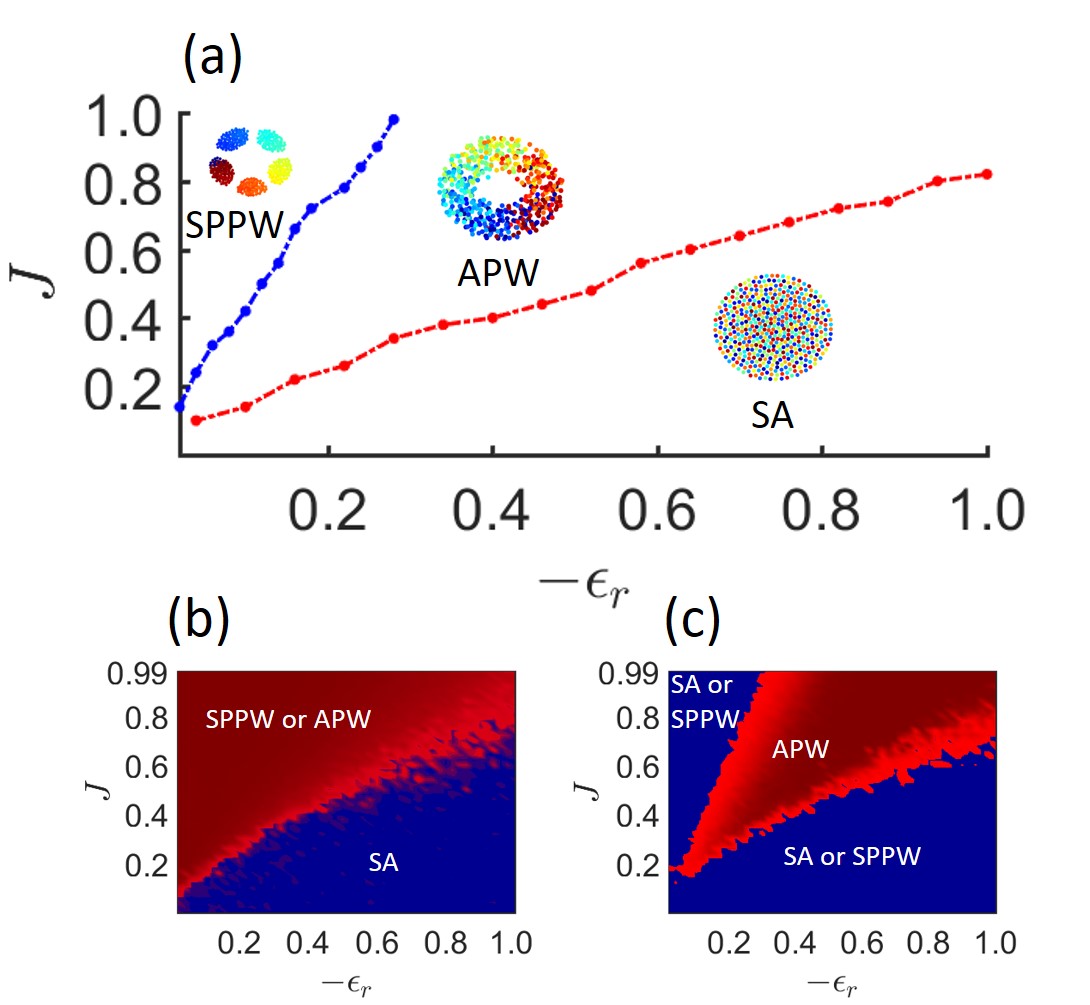}}					
	\caption{{\bf $\epsilon_r$-$J$ parameter space when $r<\delta$}: We are able to map different emergent collective states in the two-dimensional $\epsilon_r$-$J$ parameter space for a small attractive vision radius $r=10^{-5} < \delta$. We perform the numerical simulations with $N=100$ swarmalators for $500$ time units with step size $0.01$ and take last half of the data to calculate $T$ and last $10$$\%$ data points for $S$. For each point, 10 realizations are taken. In subfigure (a), the red dots are points where $S$ bifurcates from zero to non-zero value and blue dots are the points, where $T$ bifurcates from zero when $S$ is non-zero. The points have been joined by dashed line to separate the region of occurrence of SA ($S = T = 0$), APW ($S \ne 0$, $T \ne 0$), and SPPW ($S \ne 0$, $T = 0$) states. To demonstrate these bifurcations more accurately, we plot two subfigures (b) and (c). The subfigure (b) is classified based on the order parameter $S$, where the red region signifies non-zero $S$ values indicating the emergence of either SPPW or APW. The blue region contemplates the SA state with $S=0$. To distinguish between SPPW and APW states, we utilize the order parameter $T$ in subfigure (c), where $T=0$ (blue region) highlights the SA state or SPPW state. The non-zero values of $T$ (red region) reveal the manifestation of the APW state. For further information, please see the main text.}
	\label{Fig.3}
\end{figure*}

\par On the extreme opposite to the previous scenario, when the vision radius ($r$) is very small (less than the lower bound of minimal inter-particle distance, $\delta$), every swarmalator lies outside the vision range of every other swarmalators. They lose their connection with others which is required for attractive phase coupling. The presence of only repulsive phase interaction is validated by the fact that $\text{A}$ is the zero matrix now. The phase dynamics in this case is governed by the equation
\begin{equation}	
	\dot{\theta}_{i} =\frac{\epsilon_r}{N-1}\sum_{\substack{j = 1\\j \neq i}}^{N}\frac{\sin(\theta_{ij})}{|\textbf{x}_{ij}|^\gamma} \hspace{0.2cm}(\epsilon_r<0).
	\label{eq.11}
\end{equation}
Here, we observe three different asymptotic behaviors of swarmalators depending on the parameter values of $J$ and $\epsilon_r$ of Eqs.\ \eqref{eq.3} and \eqref{eq.11}.

One of them is a static state where the positions and phases of swarmalators become static and their phases are distributed over $[-\pi, \pi)$, called static asynchrony (SA) (in Fig.\ \ref{Fig.2}(b)). In the other two states, called splintered phase wave (SPPW) and active phase wave (APW), swarmalators move. For a small value of $|\epsilon_r|$, swarmalators break into clusters of distinct phases and inside each cluster, they execute small oscillations about their mean values of both position and phase. But they do not move from one cluster to another once this SPPW state is achieved (Fig.\ \ref{Fig.2}(d)). The order parameter $S$ gives nonzero value here indicating some correlation between their phase and spatial angle. If the value of $\epsilon_r$ is decreased, the SPPW pattern vanishes and swarmalators start to rotate in space and their phases execute full cycle from $-\pi$ to $\pi$ (Fig.\ \ref{Fig.2}(e)). In this APW state, $S$ also gives nonzero value which makes it impossible to distinguish these two states from each other only using $S$. Following the behavior of swarmalators in the APW state another order parameter $T$ is defined, which is the fraction of swarmalators that execute at least one full cycle in space and phase after discarding the transient. In Fig. \ref{Fig.3}, we investigate the $\epsilon_r$-$J$ parameter space to identify the regions of existence of SA, SPPW, and APW states, respectively. In the SA state, both the order parameters $S$ and $T$ give zero values. $S$ is non-zero and $T$ is zero in SPPW, whereas both non-zero in the APW state. We summarize all this information in Table \ref{tab:table2}. The stationary nature of the SA state trivially makes $T = 0$ as swarmalators are static in space, and the asynchronous behavior of their phases gives $S = 0$ which can be seen in Fig.\ \ref{Fig.3}. There is some correlation between swarmalators' spatial positions and internal phases both in the SPPW and APW states and as a result, $S$ is non-zero in both these states (see Figs.\ \ref{Fig.3}(b) and \ref{Fig.3}(c)). The inability of $S$ to distinguish these two states calls for the order parameter $T$ which is non-zero only in the APW state (see Fig.\ \ref{Fig.3}(c)).

\begin{table}
	\caption{\label{tab:table2} Different values of $S$ and $T$ are used to separate the $\epsilon_r$-$J$ space in Fig.\ \ref{Fig.3}.}
	\centerline{
		\begin{tabular}{ccccc}
			
			$S$&$T$&Emerging state\\ \hline
			$\approx 0$&$\approx 0$& Static async (SA) \\
			$\ne 0$&$\approx 0$& Splintered phase wave (SPPW) \\
			$\ne 0$&$\ne 0$& Active phase wave (APW) \\
	\end{tabular}}
\end{table}

\subsection{\label{subsec:level2}Local attractive coupling}
In the absence of negative phase coupling (i.e., $\epsilon_r = 0$), the swarmalators are only allowed to couple their phases with positive strength ($\epsilon_a$) among nearby neighbors, i.e., with other swarmalators which lie inside their range of vision. This is the case where swarmalators staying far apart from each other in space lose the connection between them to have an influence on each other's phase. For a suitable low value of vision radius $r$, the attractive adjacency matrix $\text{A}$ becomes null and both $\epsilon_a = \epsilon_r = 0$ effectively. The swarmalators are locked to their initial phases and they rearrange themselves in space following their spatial dynamics. For $J = 0$, the influence of phase on position is absent and as a result swarmalators arrange themselves inside a disc following only swarming dynamics while their phases are uniformly distributed in the range $[-\pi, \pi)$. They become static in space and phase settling into the SA state. For a non-zero value of $J$, the swarmalators in similar phase attract each other and form an annulus like structure, whose inner and outer radii depend on the values of $J$, and eventually become static in space and phase. In this state, known as static phase wave (STPW) (see Fig.\ \ref{Fig.2}(c)), spatial angle $\phi$ and phase $\theta$ of every swarmalator is perfectly correlated, which gives $S = 1$. This state is a static state and $T$ gives zero value.

\begin{figure}[hpt]
	\centerline{
		\includegraphics[scale = 0.55]{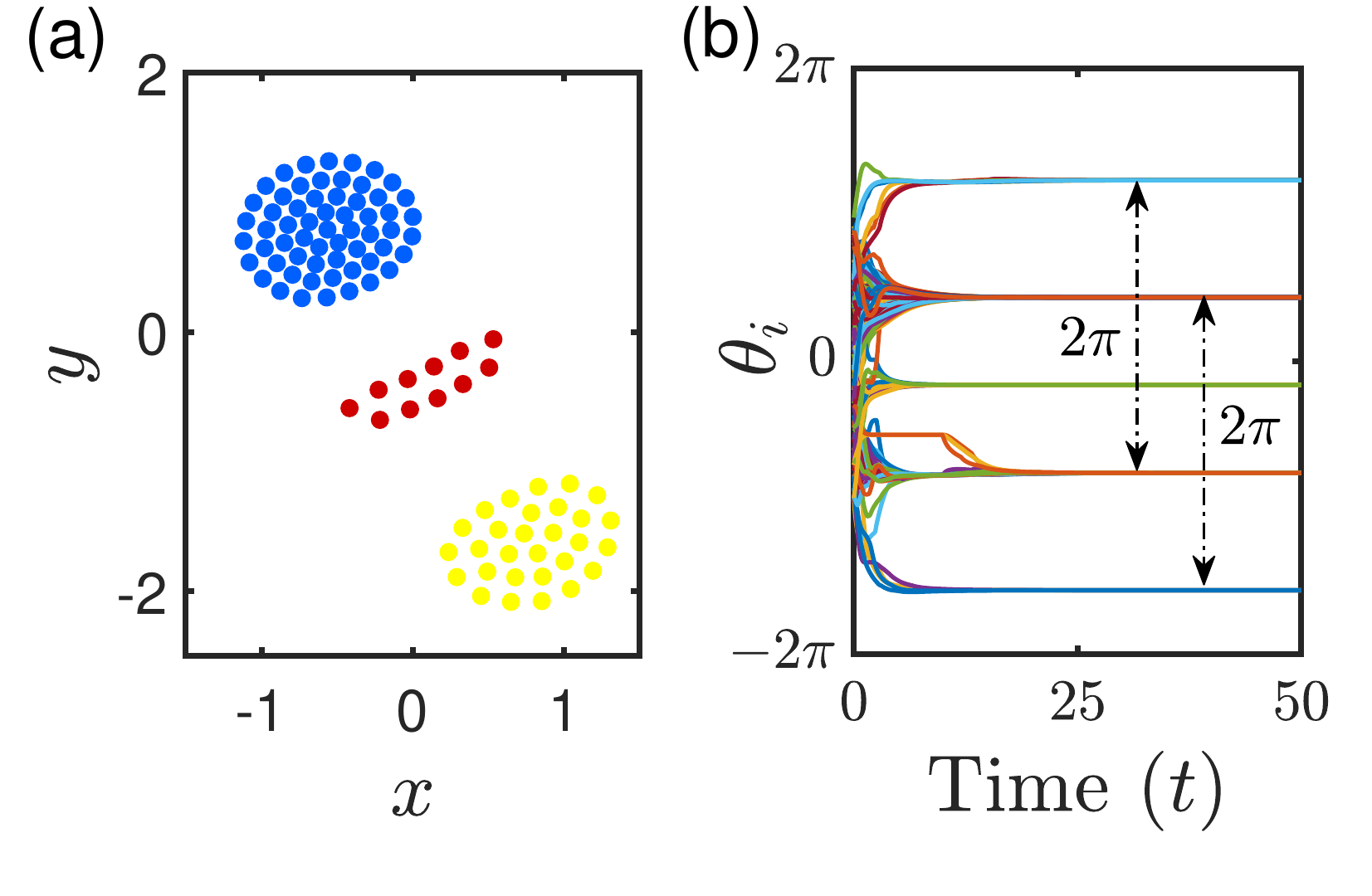}}				
	\caption{{\bf Emergence of static cluster sync state for $r = 0.3$, $J = 0.8$, and $\epsilon_a = 0.5$ in the absence of repulsive coupling}: (a) Snapshot of the swarmalators at $t = 50$ time units where the swarmalators are colored according to their phase. (b) Phase time series of the swarmalators over $t = 50$ time units. Swarmalators converge into five different stationary phases. The phase difference of $2\pi$ between two clusters represents the identical phase. Thus, the system settles down here into three distinct clusters. Evolution of a static cluster synchrony state from initial positions is shown by a video (Figure4a.gif) in the GitHub repository \cite{web_5}.}
	\label{Fig.4}
\end{figure}

When the radius of vision is increased beyond a critical value $r \approx 0.2$, swarmalators start to feel the presence of others inside their interaction range and couple their phases attractively to minimize their respective phase differences. This enhances phase similarity between nearby swarmalators. For large $r$, all their phases get fully synchronized and the emergence of static synchrony is observed. However, for a large value of $J$, the spatial attraction strength between swarmalators, given by $1 + J \cos(\theta_j - \theta_{i})$, is much bigger for the ones who are in similar phases than for those who share a relatively higher phase difference. This forces them to form groups among themselves where in each group, swarmalators' phases get totally synchronized. Between two distinct groups there is always a difference of phases and in space they are at least $r$ distance away from one another. We name this new state as {\it static cluster synchrony} (SCS), as eventually swarmalators cease to move in space and their phases become static. Note that, this formation of clusters in space happens because of the fact that swarmalators in similar phases attract themselves in space with more strength than others, even when there is global spatial attraction and repulsion among them. We validate the phase dependence of position of swarmalators in Fig.\ \ref{Fig.4}(a) by snapshot of this state where swarmalators are grouped into three clusters and they are colored according to their phases. Figure \ref{Fig.4}(b) shows their phase time series which confirms the fact that their phases become static and are divided into three groups.  The phases of swarmalators remain bounded between $-2\pi$ to $2\pi$. However this range varies depending on the choice of initial conditions. It is worth mentioning that the number of such clusters and number of swarmalators inside each cluster depend on the initial spatial position and phase distributions. The emergence of these states is independent of the values of $\epsilon_a$ when it is varied within $0$ and $1$. Generally, a natural tendency among the coupled swarmalators is to reduce their respective phase differences beyond a critical value of attractive coupling strength $\epsilon_a > 0$. However, once the clusters are formed, they remain beyond their attractive vision radius for the observed static cluster sync state. As a result, they maintain their phase differences and are never able to merge into a single cluster, irrespective of $\epsilon_a \in (0,1]$.  We include a video in the GitHub repository \cite{web_5} showing the formation of static cluster sync state from initial configurations as mentioned in Fig.\ \ref{Fig.4}.

\begin{figure}[hpt]
	\centerline{
		\includegraphics[scale = 0.5]{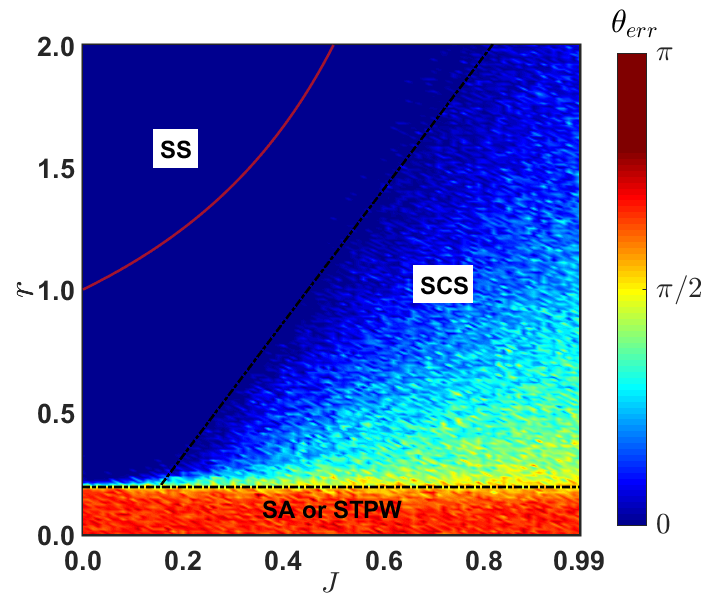}}					
	\caption{{\bf Emergent collective behavior based on the phase synchronization error in the $J$-$r$ parameter plane}: Simulation is done with $N = 100$ swarmalators for $t = 500$ time units with step size 0.01 and $\theta_{err}$ is calculated over last 10 time units. Each point is achieved using 20 realizations. We plot the analytically calculated critical curve $r = \frac{1}{1-J}$ (red) beyond which one can anticipate the occurrence of SS state. Note that this theoretically calculated curve provides a sufficient condition for the manifestation of the SS state. The black dashed lines are used to distinguish three observed fascinating states. Here $\epsilon_a = 0.5$.}
	\label{Fig.5}
\end{figure}

It is evident that whether swarmalators will end up inside a single disc (SS) or break into two or more clusters (SCS) is determined by the interplay between vision radius ($r$) and the parameter $J$. We try to approach this numerically by introducing the phase synchronization error defined by

\begin{equation}
	\theta_{err} = \Bigg \langle \frac{\sum_{i =2}^{N} |\theta_i - \theta_1|}{(N-1)} \Bigg \rangle_t,
	\label{eq.12}
\end{equation}
where $\langle \cdots \rangle_t$ stands for time average. For calculating $\theta_{err}$, we bring the phase values of swarmalators between $0$ and $2\pi$ so that swarmalators whose phase difference is an integer multiple of $2\pi$ represent the same phase. Under this scenario, our numerical simulations suggest the phase synchronization error $\theta_{err}$ takes minimum value $0$ and maximum value $\pi$. In Fig.\ \ref{Fig.5}, we plot the $J$-$r$ parameter space with the color bar indicating the phase synchronization error ($\theta_{err}$). When $r < 0.2$ (in the SA or STPW states), the value of $\theta_{err}$ is very high since the phases vary uniformly between $-\pi$ to $\pi$. For $r > 0.2$, phase synchrony starts to occur either among all units or in groups. In SS state, the synchronization error gives zero value while intermediate non-zero values of $\theta_{err}$ are observed in SCS state as synchrony is present only among the elements of each cluster but not globally. 
\par Now we investigate the sufficient condition for global synchronization of the swarmalators. It is to be noted that global synchronization occurs when two or more clusters overcome their phase differences and coincide in a unique phase. For a fixed $J $ if we gradually increase $r$, cluster synchrony states are found after the initial asynchronous SA or STPW states. Two or more clusters establish connection for phase attraction among them and merge into a single cluster by minimizing their phase differences when $r$ is further increased. The static sync state is achieved when two separate clusters merge to form a single one where their phases are fully coherent.

Let $\mathcal{C}_i$ be the set of indices of the swarmalators belonging to the $i$-th cluster for $i \in \{1,2\}$. We define 
\begin{equation}
	\textbf{x}_{c_i} = \frac{1}{|\mathcal{C}_i|} \sum_{j \in \mathcal{C}_i} \textbf{x}_{j},
	\label{eq.13}
\end{equation}
as the center of position of $i$-th cluster. When the static two-cluster state is formed, we can consider the swarmalators as a two-particle system where these two particles are positioned at the center of position of their respective clusters. Let their positions be denoted by $\textbf{x}_{c_1}$ and $\textbf{x}_{c_2}$ where $\textbf{x}_{c_1} \ne \textbf{x}_{c_2}$ and their phases be $\theta_{c_1}$ and $\theta_{c_2}$ ($\theta_{c_1} \ne \theta_{c_2}$), respectively. Since the system becomes stationary, following Eq.\ \eqref{eq.3}, we can write

\begin{equation}
	{\bf 0} = \left[ \frac{\textbf{x}_{c_2} - \textbf{x}_{c_1}}{|\textbf{x}_{c_2} - \textbf{x}_{c_1}|} (1+J\cos(\theta_{c_2} - \theta_{c_1}))  -    \frac{\textbf{x}_{c_2} - \textbf{x}_{c_1}}{{|\textbf{x}_{c_2} - \textbf{x}_{c_1}|^2}} \right],
	\label{eq.14}
\end{equation}

which gives
\begin{equation}
	|\textbf{x}_{c_2} - \textbf{x}_{c_1}| = \frac{1}{1+J\cos(\theta_{c_2} - \theta_{c_1})} \le \frac{1}{1-J}.
	\label{eq.15}
\end{equation}

So, the maximum spatial distance between the centers of the clusters is less than or equal to $\frac{1}{1-J}$. If $r$ is chosen to be greater than this value then the clusters synchronize their phases and as a result will merge to a single cluster to give the static synchrony state, i.e., $r>\frac{1}{1-J}$ is the sufficient condition for static sync state to take place which has been validated in Fig.\ \ref{Fig.5} by the red curve.

\subsection{\label{subsec:level3}Attractive-repulsive phase coupling}
We have modeled the swarmalators (Eqs.\ \eqref{eq.3} and \eqref{eq.4}) such that at short distance ($|\textbf{x}_{j}-\textbf{x}_{i}| \le r$) their phases are coupled positively and at long distance ($|\textbf{x}_{j}-\textbf{x}_{i}| > r$), they interact with negative phase coupling strength. It is easy to see, how swarmalators couple their phases depends strictly on the choice of vision radius $r$. We have already discussed the two cases (Sec.~\ref{subsec:level1}) for $r$, when it is very small and infinitely large. For small $r$, the repulsive coupling prevails over attractive coupling. When vision radius is increased, number of swarmalators inside the interaction range of each increases and attractive coupling between them takes place. For a large value of $r$, this attractive coupling completely dominates the repulsive coupling in the sense that more number of swarmalators are coupled attractively than repulsively. So, increment in the value of $r$ ensures the transition from repulsive dominance to attractive dominance in the phase coupling.

\begin{figure*}[hpt]
	\centerline{
		\includegraphics[scale = 0.65]{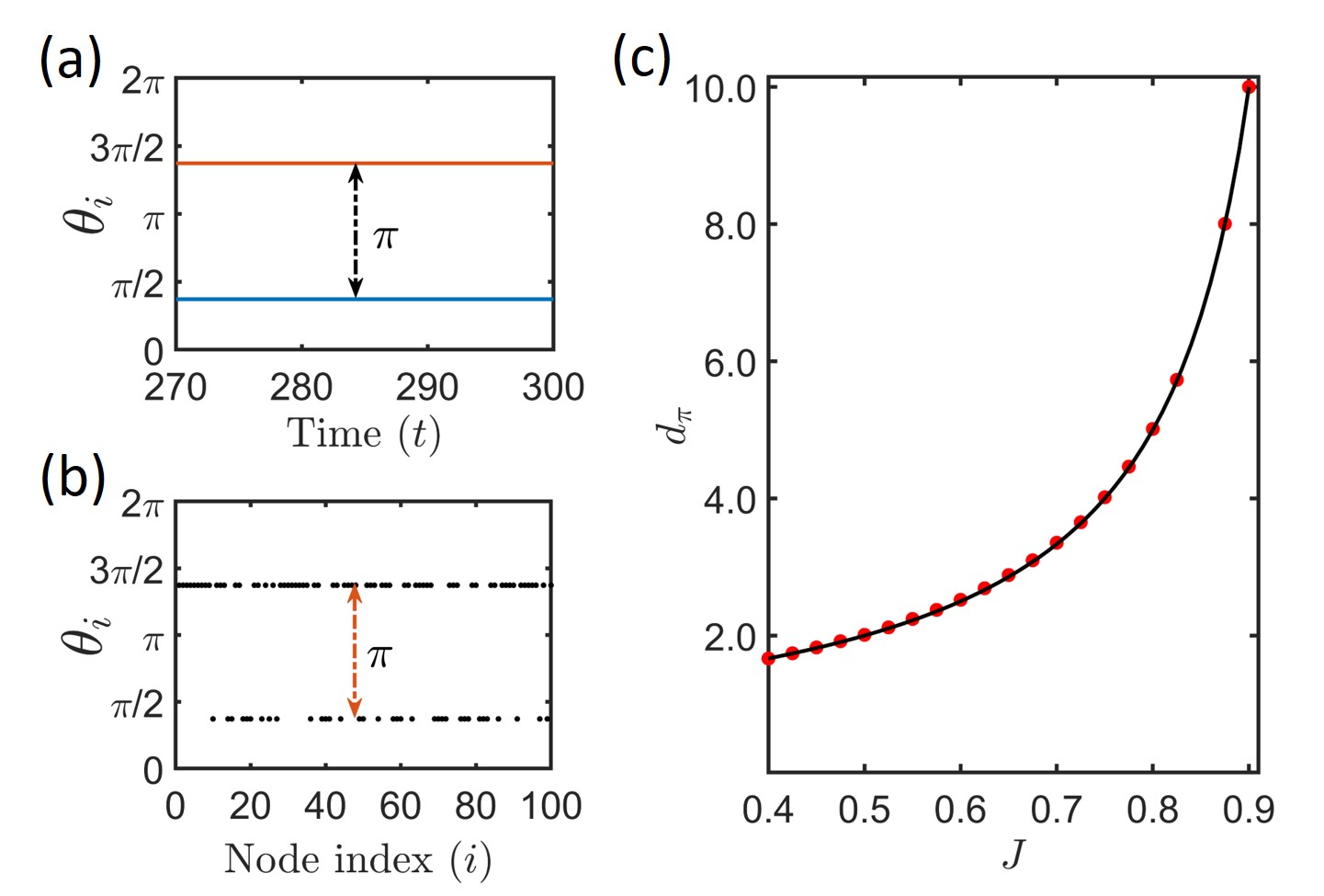}}					
	\caption{{\bf Static $\pi$ state}: (a) Time series of $N = 100$ swarmalators are shown over $t =300$ time units after transient time for $r = 0.8$ and $J = 0.5$. (b) We display the corresponding snapshot of node index ($i$) vs phase ($\theta_i$) at $t = 300$ time unit. (c) Distance between the centers of positions of the clusters, $d_\pi$ as a function of $J$ is plotted here for $r = 0.75$. There exists a critical value of $J$ depending other system parameters beyond which one can expect such SPI state. Red dots represent simulated data, while the black curve shows theoretical result given by Eq.\ \eqref{eq.16}. Evolution of the static $\pi$ state for $r = 0.8, J = 0.5, \epsilon_a = 0.5, \epsilon_r = -0.5$ is shown by a video (Figure6.gif) in the GitHub repository \cite{web_5}.}
	\label{Fig.6}
\end{figure*}

\paragraph*{Static $\pi$ state:} The co-existence of attractive and repulsive couplings in the system makes phase interaction between swarmalators complex and gives rise to richer collective behaviors. Like in the case of only attractive coupling, here also swarmalators group themselves into clusters depending on the values of $r$ and $J$. But the presence of repulsive coupling between the units of clusters forces them to maximize their phase difference and eventually the phase difference becomes $\pi$ for a stable solution. For static state, our numerical simulations assure the phase difference between the clusters is either 0 or $\pi$ (modulo $2\pi$). If clusters share a phase difference $0$ (modulo $2\pi$) then the units inside those clusters attract each other in space (note that, the spatial attraction strength is $1+J$ if phase difference $\theta_{ij} = 0$) and they form a single cluster. As a result, number of clusters is exactly two and the phase difference between them is $\pi$. We designate this new state as {\it static $\pi$ state} (SPI) since swarmalators become stationary in phase and space eventually. As the swarmalators inside each cluster are fully synchronized but not with the ones inside the other cluster, $R$ gives a non-zero value and is less than $1$. The order parameter $S$ also gives an intermediate non-zero value between $0$ and $1$ in this state, demonstrating the presence of correlation between swarmalators' spatial angles and phases. In Fig.\ \ref{Fig.6}(a), the temporal behavior of 100 swarmalators is shown, where the phases become static and are clustered maintaining exactly a phase-difference $\pi$. At a particular time instant after the transient time, their phases are plotted versus their respective node indices in Fig.\ \ref{Fig.6}(b), which validates the clustering of swarmalators in two groups.  In Fig.\ \ref{Fig.6}(a)-(b), the phases of swarmalators are plotted after taking modulo $2\pi$ so that swarmalators at a phase difference of integer multiple of $2\pi$ represent the same phase. A video showing the origination of static $\pi$ state is included for visual understating as highlighted in Fig.\ \ref{Fig.6}.

In the SPI state, we can consider the swarmalators as a two-particle system where they are positioned at the center of position of their respective cluster having a phase difference $\pm \pi$. Let their positions be denoted by $\textbf{x}_{c_1}$ and $\textbf{x}_{c_2}$, where $\textbf{x}_{c_1} \ne \textbf{x}_{c_2}$. Then from Eq.\ \eqref{eq.14}, we can write

%

\begin{equation}
	|\textbf{x}_{c_2} - \textbf{x}_{c_1}| = \frac{1}{(1-J)} = d_\pi,
	\label{eq.16}
\end{equation}

since $\theta_{c_2} - \theta_{c_1} = \pm \pi$ and $\textbf{x}_{c_2} - \textbf{x}_{c_1} \ne {\bf 0}$.

Equation\ \eqref{eq.16} gives the distance between the centers of position of the clusters ($d_\pi$) in SPI state which increases when we increase the value of $J$. Figure \ref{Fig.6}(c) shows the change of $d_\pi$ with increment of $J$ in the interval where SPI state is achieved.

\begin{figure*}[hpt]
	\centerline{
		\includegraphics[scale = 0.55]{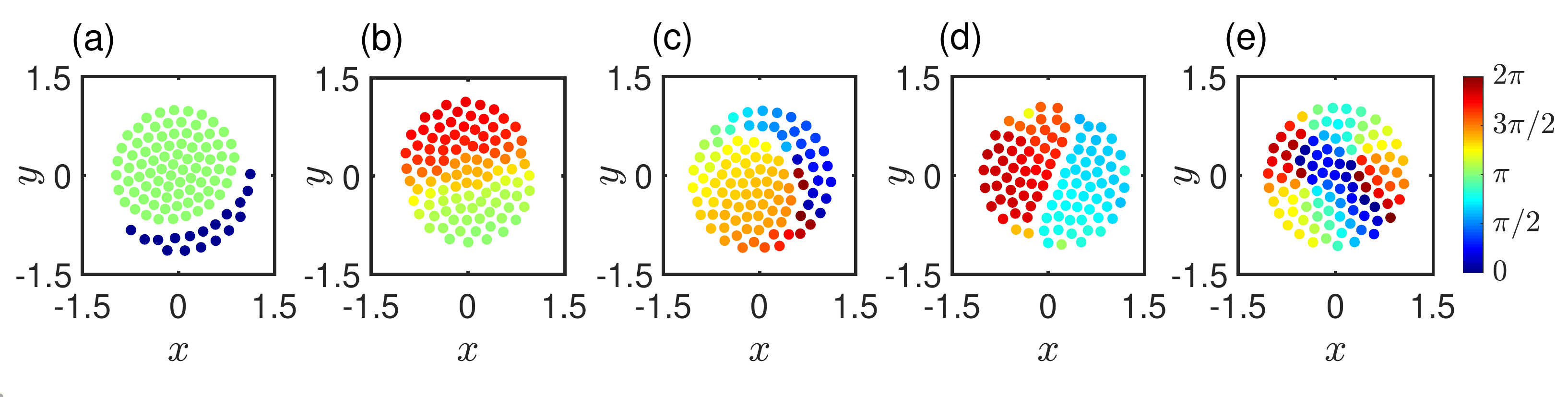}}
	\caption{{\bf Snapshots of mixed phase wave states}: Simulations are performed for $N = 100$ swarmalators over 500 time units with step size 0.01 and $J = 0.1$, and phases are changing over $0$ to $2\pi$. Attractive mixed phase wave is observed for (a) $\epsilon_a = 1.0$, $\epsilon_r = -0.1$, $r = 0.21$, (b) $\epsilon_a = 0.5$, $\epsilon_r = -0.5$, $r = 1.38$, and (c) $\epsilon_a = 0.1$, $\epsilon_r = -1.0$, $r = 1.83$. Swarmalators show the tendency of forming clusters with the ones in nearby phases, but can not achieve separate cluster formation for small values of $J$. Repulsive mixed phase wave emerges for (d) $\epsilon_a = 0.5$, $\epsilon_r = -0.5$, $r = 0.21$, and (e) $\epsilon_a = 0.1$, $\epsilon_r = -1.0$, $r = 0.36$. Repulsive coupling is dominant on attractive coupling for sufficiently small $r$ and when $\epsilon_a \ll |\epsilon_r|$. This state is like merging of separate clusters in splintered phase wave state. Evolution of these states corresponding to (a)-(e) are shown by five videos (Figure7a.gif-Figure7e.gif) in the GitHub repository \cite{web_5}.}
	\label{Fig.7}					
\end{figure*}

\paragraph*{Mixed phase wave state:} Significantly, for a small value of $J$, swarmalators accumulate themselves in space with the ones in nearby phase, but they do not form distinct clusters. The positive phase coupling induces minimization of phase difference between spatially nearby swarmalators, and at the same time, negative coupling ensures that there is some phase difference between swarmalators who lie outside the vision range. They form a deformed disc like structure, where swarmalators move inside it maintaining phase similarity with nearby units. This active state is different from the two previously mentioned active states (SPPW and APW states), as swarmalators neither break into disjoint clusters nor execute full cycle in space and phase. In this state, the swarmalators are neither completely phase synchronized nor fully desynchronized, rather they show an intermediate behavior. We name this state as {\it mixed phase wave} (MPW) state. In Fig.\ \ref{Fig.7}, snapshots of this state are shown for different values of parameters.  Like in the previous cases, again we plot the swarmalators after bringing their phases in the interval $[0, 2\pi]$. Corresponding to Fig.\ \ref{Fig.7}, five videos showing the evolution of mixed phase wave states are incorporated in the GitHub repository \cite{web_5}. This kind of behavior can be seen when swarmalators are on the verge of forming splintered phase wave or static $\pi$ but they can not attract other swarmalators in nearby phases with enough strength to make disjoint clusters. We classify this mixed phase wave state into two categories. One is repulsive mixed phase wave (RMPW) state (where repulsive coupling dominates and qualitatively similar to SPPW) and the other is attractive mixed phase wave (AMPW) state (which is dominated by attractive coupling and qualitatively similar to SPI). In both the AMPW and RMPW states, $S$ is non-zero indicating the presence of some correlation among swarmalators' phases and spatial angles. In the attractive coupling influenced AMPW state, the order parameter $R$ gives a non-zero value showing some coherence among swarmalators' phases. The value of $R$ is zero in the repulsive coupling dominated RMPW state. So by using the value of $R$, we separate these two states from each other.

\begin{figure*}[hpt]
	\centerline{
		\includegraphics[width = 16cm,height = 12cm]{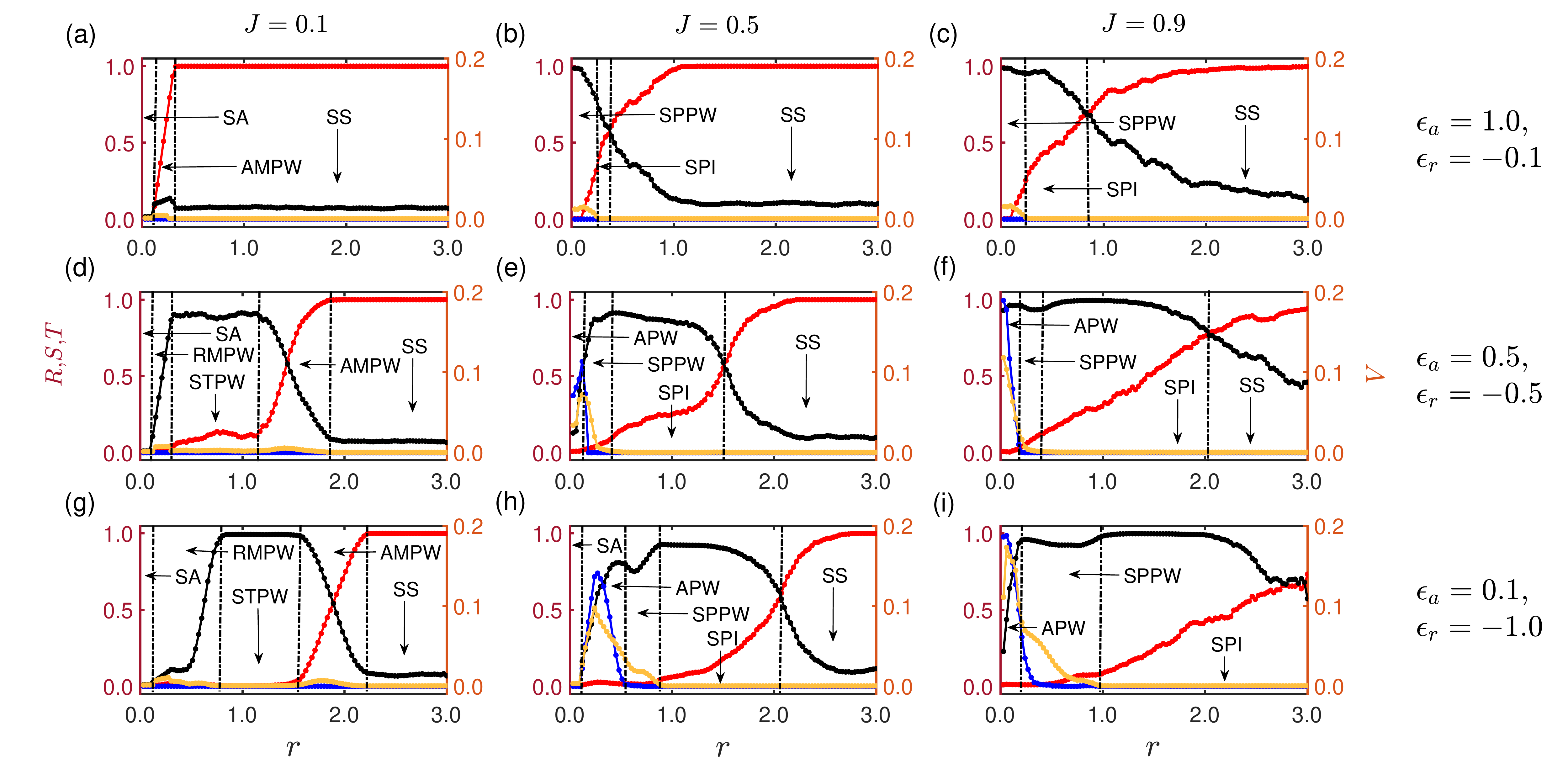}}
	\caption{{\bf Transition of states with varying $r$ for different choices of $J$, $\epsilon_a$, and $\epsilon_r$}: Behavior of order parameters $R$ (red), $S$ (black), $T$ (blue), and $V$ (yellow) with varying $r$ are shown. $\epsilon_a = 1.0$ and $\epsilon_r = -0.1$ in (a)-(c), $\epsilon_a = 0.5$, and $\epsilon_r = -0.5$ in (d)-(f), and $\epsilon_a = 0.1$ and $\epsilon_r = -1.0$ in (g)-(i). $J = 0.1$ in (a), (d), (g), $J = 0.5$ in (b), (e), (h), and $J = 0.9$ in (c), (f), (i). Simulations are done for $N = 100$ swarmalators over 500 time units with integration step size 0.01. Last 10$\%$ of the data are chosen for calculating $R$ and $S$, whereas last 50$\%$ data are used to calculate $T$. $V$ is calculated by taking time average over last 50 time units. For each point, a mean of 20 realizations is taken. $R$, $S$, and $T$ are plotted along left y axis and $V$ is plotted along the right y axis.}
	\label{Fig.8}					
\end{figure*}

\par To separate the stationary states from the non-stationary ones, we measure the mean velocity $V$ defined as,

\begin{equation}
	V = \Bigg \langle\frac{1}{N} \sum_{i=1}^{N} \sqrt{\dot{x}_i^2 + \dot{y}_i^2 + \dot{\theta}_i^2} \Bigg \rangle_t,
	\label{eq.17}
\end{equation}
where the time average is taken after discarding the transients. A finite non-zero value of the mean velocity $V$ suggests that swarmalators move in space, and their phases evolve within the interval $[0,2\pi)$, while in stationary states $V$ is zero. Depending on the values of $r$, $J$, $\epsilon_a$, and $\epsilon_r$, the swarmalators exhibit different long-term states. In Fig.\ \ref{Fig.8}, we vary $r$ over the range $(0,3]$ keeping the other three parameters fixed and examine the transition of states. {We consider three different scenarios for three different values of $J = 0.1, 0.5$, and $0.9$. Depending on the values of $\epsilon_a$ and $\epsilon_r$, three different cases can be implemented:} 
\paragraph*{Case-1:} $\epsilon_a > |\epsilon_r|$. This condition implies that the attractive coupling strength between the swarmalators within their vision range is larger than the repulsive coupling strength with outer swarmalators.
\paragraph*{Case-2:} $\epsilon_a = |\epsilon_r|$. The swarmalators interact with same strength with others, both inside and outside their vision range but with opposite sign.
\paragraph*{Case-3:} $\epsilon_a < |\epsilon_r|$. Here, the attractive coupling strength between swarmalators within their vision range is smaller than the repulsive coupling strength for interaction outside it.

\begin{table*}[ht]
	\caption{\label{tab:table3}This table shows how the emerging states of swarmalator system are identified with the order parameters $R$, $S$, $T$, and $V$. }
	\centerline{
		\begin{tabular}{ccccc}
			
			$R$&$S$&$T$&$V$&Emerging state\\ \hline
			$\approx$ 1&$0 < S < R$&$\approx 0$ &$\approx 0$& Static sync (SS) \\
			$\approx 0$&$\approx 0$&$\approx 0$&$\approx 0$&Static async (SA)\\
			$\approx 0$&$\approx 1$&$\approx 0$&$\approx 0$&Static phase wave (STPW)$^{a}$ 
			\\
			$R <1 (\ne 0)$&$S \ne 0$&$\approx 0$&$\approx 0$&Static $\pi$ (SPI)\\
			$\approx 0$&$\ne 0$ &$\approx 0$&$ \ne 0$ &Splintered phase wave (SPPW)\\
			& & & &or Repulsive mixed phase wave (RMPW)$^b$ 
			\\
			$\approx 0$&$\ne 0$ &$\ne 0$&$ \ne 0$ &Active phase wave (APW)\\
			$\ne 0$&$\ne 0$ &$\approx 0$&$ \ne 0$ &Attractive mixed phase wave (AMPW)\\
	\end{tabular}}
	
	$^{a}${In Fig.\ \ref{Fig.8}(d) for STPW state $R$ and $S$ are slightly deviated form $0$ and $1$, respectively due to the deformed structure.
		$^{b}${RMPW state is qualitatively similar to the SPPW state when they are studied with these four order parameters. RMPW state takes place when $J$ is not large enough for formation of clusters. If we increase the value of $J$ keeping the other parameters fixed, evolution of SPPW state is seen from the RMPW state}.}
\end{table*}

\par We use different order parameters to classify various collective patterns and briefly encapsulate all possible emergent states in Table \ref{tab:table3}. For the top row of Fig.\ \ref{Fig.8}, attractive phase coupling strength $\epsilon_a$ is chosen larger in modulus than the repulsive coupling strength $\epsilon_r$. In the middle row, these two are same in modulus, while $\epsilon_a < |\epsilon_r|$ for the bottom row. The value of $J$ is same in every column and $J = 0.1$, $0.5$, and $0.9$ for column 1, column 2, and column 3, respectively. In Fig.\ \ref{Fig.8}(a), $\epsilon_a = 1.0$ is bigger than the modulus of $\epsilon_r = -0.1$. Very small $r$ ensures phase coupling is repulsive and as a result SA is seen. Since $\epsilon_a$ is very larger than $|\epsilon_r|$, synchrony is achieved quiet fast while varying $r$. Before achieving complete synchrony in SS state, over a small range of values of $r$, clustering tendency is seen, but for small value of $J = 0.1$, the swarmalators are unable to break into distinct clusters and as a result AMPW state is found (see Fig.\ \ref{Fig.7}(a)). The possible route of SS state by changing the vision radius $r$ is 
\begin{equation}
	\text{SA} \rightarrow \text{AMPW} \rightarrow \text{SS}.
\end{equation}
When $J$ is increased to 0.5 in Fig.\ \ref{Fig.8}(b), emergence of SPI state is seen due to the tendency of swarmalators to group themselves. With increasing $r$, the transition of states can be seen as
\begin{equation}
	\text{SPPW} \rightarrow \text{SPI} \rightarrow \text{SS}.
\end{equation} 
Qualitative same behaviors of the order parameters are seen in Fig.\ \ref{Fig.8}(c) where $J$ is further increased to 0.9. 
\par The absolute values of $\epsilon_a$ and $\epsilon_r$ are same in Fig.\ \ref{Fig.8}(d)-\ref{Fig.8}(f). When $J = 0.1$, the swarmalators stay nearby with swarmalators in similar phases but fail to make disjoint clusters. In Fig.\ \ref{Fig.8}(d), when $r$ is very small, $\epsilon_r$ dominates the phase coupling and swarmalators are found to settle into SA state. When $r$ is increased, effect of $\epsilon_a$ starts to take place but still repulsive coupling continue to dominate. Emergence of RMPW (see Fig.\ \ref{Fig.7}(d)) is seen for a small range of $r$. This state is analogous to the SPPW state, the only difference, $J$ is not large enough for the swarmalators to splinter into disjoint clusters. Further, increasing the value of $r$, a situation is achieved, where the effect of $\epsilon_a$ and $\epsilon_r$ neutralizes each other, and the swarmalators settle in the STPW state. In some cases, this STPW state is formed in a slightly deformed annular structure and, as a result, the correlation between phases and spatial angles of the swarmalators deteriorates little. This is the reason why $S$ is not exactly 1 here, otherwise $S=1$ in the STPW state. Moving to the right with increasing $r$, phase attraction starts to dominate the phase repulsion. AMPW state is noticed which is close to the SPI state (found for  bigger $J$ values). Here, $J = 0.1$ being small, the swarmalators can not break into separate clusters, but they coexist (see Fig.\ \ref{Fig.7}(b)). Finally, for reasonable larger $r$ values where the phase attraction dominates, SS state emerges. So, the transition which is seen in Fig.\ \ref{Fig.8}(d) is 
\begin{equation}
	\text{SA} \rightarrow \text{RMPW} \rightarrow \text{STPW} \rightarrow \text{AMPW} \rightarrow \text{SS}.
\end{equation}
When $J$ is increased to 0.5, formation of clusters happens which was not seen in the case of $J = 0.1$. For very small $r$, the swarmalators are effectively phase coupled only with $\epsilon_r$ and as a result, APW is found which is shown in Fig.\ \ref{Fig.8}(e). Here, due to the ability to break into clusters, SPPW state is seen in place of RMPW in Fig.\ \ref{Fig.8}(d). Increment in the value of $r$ increases phase attraction among the swarmalators and SPI state is found to occur before the ultimate SS state. The route to achieve SS state in Fig.\ \ref{Fig.8}(e) is
\begin{equation}
	\text{APW} \rightarrow \text{SPPW} \rightarrow \text{SPI} \rightarrow \text{SS}.
\end{equation}
For $J = 0.9$ in Fig.\ \ref{Fig.8}(f), the transitions are same as Fig.\ \ref{Fig.8}(e).

\par In the bottom row of Fig.\ \ref{Fig.8} $\epsilon_a = 0.1$ is relatively small than $|\epsilon_r| = 1.0$. For $J = 0.1$ in Fig.\ \ref{Fig.8}(g), the evolution of states takes place in the same manner as in Fig.\ \ref{Fig.8}(d). In Fig.\ \ref{Fig.8}(h), the route for SS state is
\begin{equation}
	\text{SA} \rightarrow \text{APW} \rightarrow \text{SPPW} \rightarrow \text{SPI} \rightarrow SS,
\end{equation}
where $J = 0.5$. With $J = 0.9$ in Fig.\ \ref{Fig.8}(i), we can see that the SPI state exists over a long range of $r$ and SS is yet not achieved.

\subsection{\label{sec:level8} Effect of negative $J$}

The parameter $J$ which stands for the effect of phase similarity on spatial attraction, plays a significant role in determining the asymptotic state of the swarmalator system defined by Eqs.\ \eqref{eq.3} and \eqref{eq.4}. So far the results in our work are accomplished by considering $J$ to be strictly positive. A positive value of $J$ makes sure that the spatial attraction strength between nearby phased swarmalators is increased and as a result they stay nearby to each other. A negative value of the parameter $J$ completely alters the scenario. Now swarmalators which are in opposite phases attract each other spatially with more strength than those which are in similar phases. This reverse phenomena changes the complexion of the swarmalator system altogether. At the same time it increases the complexity of our model. To capture the effect when a negative $J$ is considered, we study the case of local attractive coupling with $J<0$. Also to show the generic nature of our model for the exponents $\alpha$, $\beta$, and $\gamma$ (as long as $\alpha < \beta$ holds), here we consider $\alpha = 0$ (note that, all the studies so far has been done with $\alpha=1$). Now, with these assumptions we write down the governing equations as

\begin{equation}
	\label{eq.39}
	\dot{\textbf{x}}_{i} = \frac{1}{N-1} \sum_{\substack{j = 1\\j \neq i}}^{N}\left[ \textbf{x}_{ij} (1+J\cos(\theta_{ij}))  -    \frac{\textbf{x}_{ij}}{{|\textbf{x}_{ij}|}^2} \right],
\end{equation}

\begin{equation}	
	\label{eq.40}
	\dot{\theta}_{i} = \frac{\epsilon_a}{N_{i}}\sum_{\substack{j = 1\\j \neq i}}^{N} \text{A}_{ij}\frac{\sin(\theta_{ij})}{|\textbf{x}_{ij}|},
\end{equation}
where $\text{A}$ is the adjacency matrix for local attractive phase coupling described in Sec.\ \ref{sec:level2}. This model inherits contrasting spatial and phase interaction. The local attractive phase coupling among the swarmalators minimizes the phase difference among spatially nearby ones. On the other hand, swarmalators which are in nearby phases reduce the spatial attraction among them (since $J$ is negative). 

\begin{figure*}[hpt]
	\centerline{
		\includegraphics[scale = 0.48]{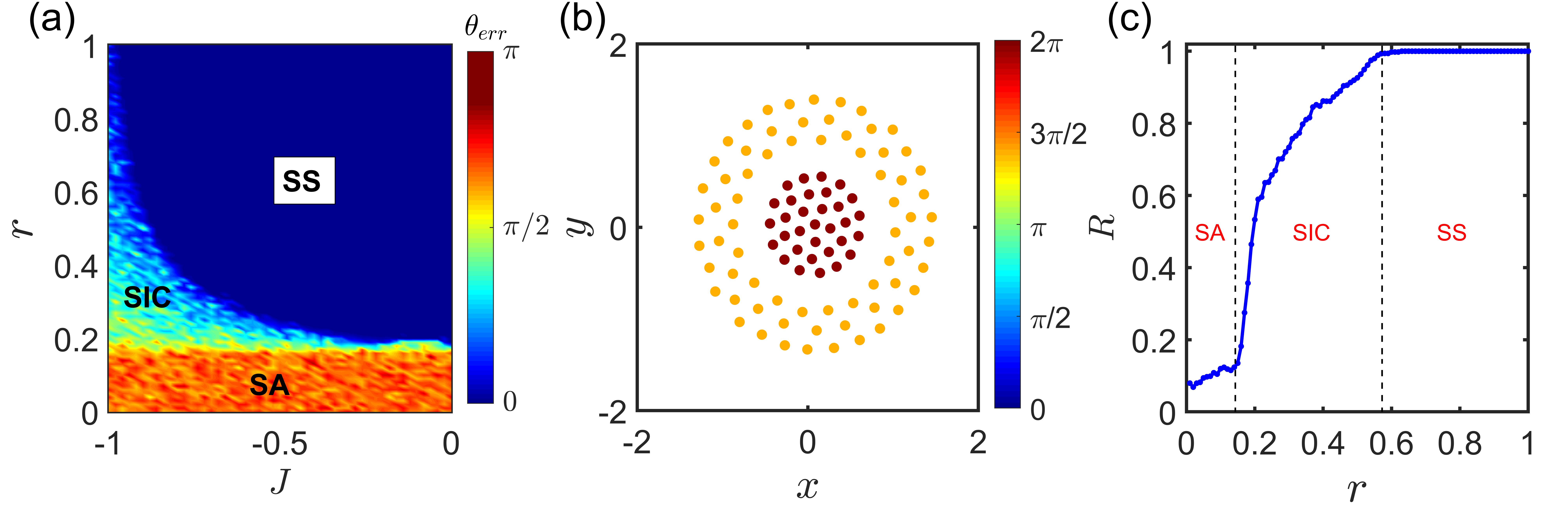}}					
	\caption{{\bf Emerging asymptotic states with local attractive phase coupling when $J<0$}: The emergence of static async (SA), static inscribed cluster (SIC), and static sync (SS) is seen by considering negative $J$ under local attractive phase coupling. (a) The $J$-$r$ parameter plane is plotted based on the values of phase synchronization error $\theta_{err}$. It is maximum in the SA state and minimum in the SS state, whereas it gives intermediate values in the SIC state. (b) Snapshot of the SIC state at $t=500$ time unit for $J=-0.8$ and $r=0.4$. Swarmalators are colored according to their phases and it is seen that phases are divided into two clusters. Swarmalators whose phases are synchronized, all lie either within an annulus or a disc where the disc is inscribed within the annulus. (c) The evolution of order parameter $R$ is shown varying the value of $r$ from $0$ to $1$ for a fixed $J$ ($=-0.8$). The transition from SA state to SS state via the SIC state is seen with increasing $r$. For $R \approx 0$, the SA state is perceived. While for $R \approx 1$, the system settles down to the SS state. For any intermediate non-zero values of $R$ and $\theta_{err}$, we confirm the emergence of the SIC state by plotting the snapshot at a specific time after the initial transient.}
	\label{Fig.11}
\end{figure*}

When the vision radius $r$ is small ($r<0.2$), swarmalators can not feel the presence of one another inside their vision range and their phases remain decoupled. But negative value of $J$ induces disorder in the system where swarmalators' phases are totally desynchronized and they become stationary both in phase and space. In this static async (SA) state, the value of order parameter $R$ is close to zero. The phase synchronization error $\theta_{err}$ defined by eq.\ \eqref{eq.12} is very large (see Fig.\ \ref{Fig.11}(a)). Keeping the value of $J$ fixed when we increase the value of $r$, swarmalators start to couple their phases with spatially nearby ones and minimize their phase difference. We observe the emergence of a new pattern where swarmalators' phases are settle in two different clusters. Contrary to the static cluster state found with a positive value of $J$ for local attractive coupling in subsection~\ref{subsec:level2}, here swarmalators do not form separate clusters in the two-dimensional plane. Rather they settle themselves inside an annulus and a disc where the disc remains inscribed within the annulus. Inside both of these annulus and disc swarmalators are totally phase synchronized but they maintain a phase difference between them. Swarmalators become static after transient period both in space and phase. Following the nature of this state, we name it as the {\it static inscribed cluster} (SIC) state. In Fig.\ \ref{Fig.11}(b) we have shown the snapshot of this state in the two-dimensional plane taking $J=-0.8$ and $r = 0.4$. The order parameter $R$ gives intermediate values between $0$ and $1$ due to the presence of phase synchrony among the units of each cluster. The value of $\theta_{err}$ also decreases for the same reason compared to the SA state (see Fig.\ \ref{Fig.11}(a)). This SIC state loses its stability when we further increase the value of $r$. A large value of $r$ ensures the dominance of attractive phase coupling among all the swarmalators. As a result globally phase synchronized static sync (SS) state emerges. $R$ gives the maximum value whereas $\theta_{err}$ acquires the minimum value in this stationary state.

In Fig.\ \ref{Fig.11}(a) we have plotted the $J$-$r$ parameter space based on the values of phase synchronization error $\theta_{err}$. It is seen that $\theta_{err}$ is very large in the SA state and is almost zero in the SS state. In the SIC state it gives intermediate values between $0$ and $\pi$. To observe the transition of these states, we have plotted the order parameter $R$ against increasing $r$ in Fig.\ \ref{Fig.11}(c) where a particular value of $J$ ($=-0.8$) has been chosen. It is seen that for small $r$ ($<0.2$) in the SA state the value of $R$ is near $0$. When $r$ is increased beyond $0.2$, phase coherence among units of each clusters starts to take place in the SIC state and as a result the value of $R$ increases. Another critical value of $r$ ($\approx 0.58$) is found where this SIC state loses its stability and emergence of SS state is seen beyond that.

\subsection{\label{sec:level7}Effect of phase dynamics on spatial repulsion}

\par Till now, we have considered $\text{F}_{\text{rep}} = 1$, i.e., when there is no effect of phase similarity on spatial repulsion. Here, we investigate the scenario when swarmalators' phases affect the spatial repulsion among them. We specifically choose $\text{F}_{\text{rep}}(\theta) = 1 - K\cos(\theta)$ so that the spatial repulsion among nearby phase swarmalators is reduced when $K>0$. We consider $0<K<1$ which keeps the function $\text{F}_{\text{rep}}$ strictly positive (If $K>1$, then depending on the value of $\theta_j - \theta_i$ the function $\text{F}_{\text{rep}}$ can take negative value, which makes $\text{I}_\text{rep}$ attractive in nature). It is worth mentioning here that Theorem \eqref{Theo:1} holds for any choices of the function $\text{F}_{\text{att}}$ and $\text{F}_{\text{rep}}$ as long as they are even and bounded. In this case, $\text{F}_{\text{att}}(\theta) = 1 + J\cos(\theta)$ and $\text{F}_{\text{rep}}(\theta) = 1 - K\cos(\theta)$ are both even functions of their arguments and are also bounded. So we can safely say that Theorem \eqref{Theo:1} holds with these choices of functions which eliminates the inter-particle collision among the swarmalators. This extra interaction term in the swarmalator model can establish novel states for position aggregation and phase synchronization especially in the presence of competitive phase interaction, which requires a sincere inspection of the model. For our purpose we only consider such effect of phase similarity on spatial repulsion when the vision radius is very small. In that case our model with the choice of $\text{F}_{\text{rep}}$ can be written as
\begin{equation}
	\label{eq.26}
	\begin{split}
	\dot{\textbf{x}}_{i} = \frac{1}{N-1} \sum_{\substack{j = 1\\j \neq i}}^{N} \bigg[ \frac{\textbf{x}_{j}-\textbf{x}_{i}}{|\textbf{x}_{j}-\textbf{x}_{i}|} (1+J\cos(\theta_{j}-\theta_{i}))   \\    -\frac{\textbf{x}_{j}-\textbf{x}_{i}}{{|\textbf{x}_{j}-\textbf{x}_{i}|}^2}(1-K\cos(\theta_{j}-\theta_{i})) \bigg],
	\end{split}
\end{equation}

\begin{equation}	
	\label{eq.27}
	\dot{\theta}_{i} =  \frac{\epsilon_r}{N-1}\sum_{\substack{j = 1\\j \neq i}}^{N} \frac{\sin(\theta_{j}-\theta_{i})}{|\textbf{x}_{j}-\textbf{x}_{i}|}.
\end{equation}

\begin{figure*}[hpt]
	\centerline{
		\includegraphics[scale = 0.45]{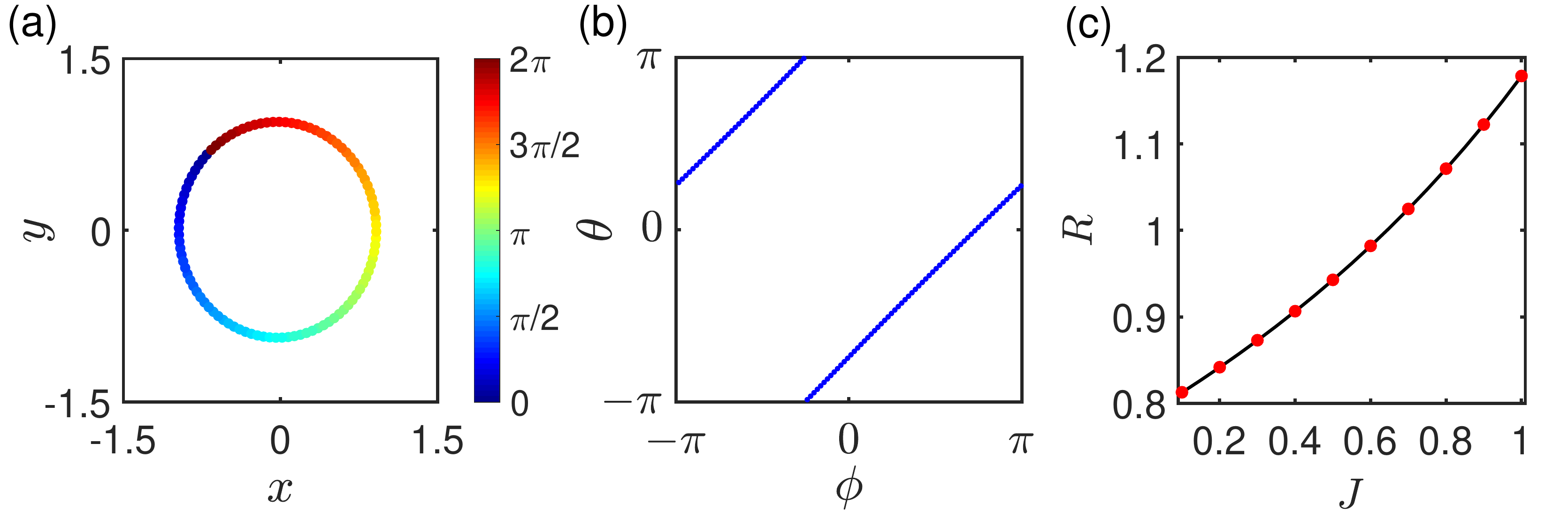}}					
	\caption{{\bf Static ring phase wave state}: We do simulations with $N=100$ swarmalators for $J=0.5$, $K=1.0$, and $\epsilon_r=-0.1$. (a) Position of swarmalators on a ring centered around the origin at $t=150$ time unit. (b) We plot the phases ($\theta$) of the swarmalators against their spatial angles ($\phi$) where perfect correlation among them is found. (c) The radius ($R$) of the ring phase wave state is calculated numerically for several values $J$ while keeping $K$ fixed at $1.0$. Red dots are the numerically calculated values of $r$. The black curve represents analytical expression given by Eq.\ \eqref{eq.38} for $K=1.0$. It is seen that, our numerical and analytical results match satisfactorily well.}
	\label{Fig.10}
\end{figure*}

Note that, here we have expressed $\textbf{x}_{ij}$ and $\theta_{ij}$ as $\textbf{x}_{j}-\textbf{x}_{i}$ and $\theta_{j}-\theta_{i}$, respectively, which is simpler to deal with while carrying out mathematical calculation. It is found that, for certain values of $J$, $K$, and $\epsilon_r$ the swarmalators organize themselves on a ring where their phases ($\theta_i$) are perfectly correlated with their spatial angles ($\phi_i$). This state is a stationary state, where after the transient period, swarmalators become static in phase as well as in the spatial positions, and we call this state as the static ring phase wave state. This static state was reported in \cite{o2018ring} for $\alpha=0$ and $\gamma=2$. In Fig.\ \ref{Fig.10}(a) swarmalators position on a ring centered around the origin is shown where they are colored according to their phases. In the static ring phase wave state, the spatial angle and phase of the swarmalators are perfectly correlated. Figure \ref{Fig.10}(b) highlights the correlation between swarmalators' spatial angles and phases. The position and phase of the $k$-th swarmalator in this state can be written as

\begin{equation}
	\textbf{x}_k = R\cos(2\pi k / N) \hat{i} + R\sin(2\pi k/N) \hat{j},
	\label{eq.28}
\end{equation}

\begin{equation}
	\theta_k = 2\pi k/N + C,
	\label{eq.29}
\end{equation}
where $R$ is the radius of the ring state, $\hat{i}$ and $\hat{j}$ are unit vectors in $x$ and $y$ directions, respectively, and the constant $C$ depends on the initial conditions. The radius $R$ can be calculated analytically. The structure of static ring phase wave state leads us to use convenient complex notation where the two-dimensional vector $\textbf{x}_{k} = (x_k,y_k)$ is identified as a point in the complex plane $z_k = (x_k,y_k)$. To calculate the radius $R$, we consider a general model of swarmalators \cite{o2018ring}

\begin{equation}
    \begin{split}
	\dot{z_i} = \frac{1}{N-1}\sum_{j=1}^{N} \big[ f(|z_i-z_j|^2)(z_i-z_j) \\+h(|z_i-z_j|^2)(z_i-z_j)\cos(\theta_i-\theta_j)\big],
	\label{eq.30}
	\end{split}
\end{equation}	

\begin{equation}
	\dot{\theta_i} = \frac{1}{N-1}\sum_{j=1}^{N} \sin(\theta_i-\theta_j) g(|z_i-z_j|^2).
	\label{eq.31}
\end{equation}

The swarmalator model defined by Eqs.\ \eqref{eq.26} and \eqref{eq.27} corresponds to Eqs.\ \eqref{eq.30} and \eqref{eq.31} when we specifically choose 

\begin{equation}
	f(r) = \frac{1}{r} - \frac{1}{\sqrt{r}}; \hspace{10pt} h(r) = -\frac{K}{r} - \frac{J}{\sqrt{r}}; \hspace{10pt} g(r) = -\frac{\epsilon_r}{\sqrt{r}}.
	\label{eq.32}
\end{equation}

In complex plane the ring phase wave state is given by

\begin{equation}
\begin{split}
	z_k = R u^k, \hspace{10pt} \text{where} \hspace{10pt} u= exp(2\pi l / N), \hspace{10pt} \\ \theta_k = 2 \pi k /N + C \hspace{10pt} (l = \sqrt{-1}).
\end{split}
\end{equation}

Although Eq.\ \eqref{eq.31} is satisfied by any values of $R$, Eq.\ \eqref{eq.30} is satisfied only if 

\begin{equation}
\begin{split}
	\sum_{i \ne 0} \big[ f(R^2|1-u^i|^2)(1-u^i) \\ + h(R^2|1-u^i|^2)(1-u^i) \cos(2\pi i/N) \big]= 0 .
	\label{eq.33}
\end{split}
\end{equation}

For the choices of the functions $f$, $h$, and $r$ given by Eq.\ \eqref{eq.32}, we get an expression for $R$ from Eq.\ \eqref{eq.33} using the identities

\begin{align}
	\sum_{i \ne 0} \frac{1}{1-u^i} &= \frac{N-1}{2}, 
\end{align}

\begin{equation}
	\sum_{i \ne 0} \frac{u^i+u^{-i}}{1-u^{-i}} = -1,
\end{equation}

\begin{equation}
	\sum_{i \ne 0} \frac{1-u^i}{|1-u^i|} = \frac{\sin(\pi/N)}{1-\cos(\pi/N)} = a,
\end{equation}

\begin{equation}
	\sum_{i \ne 0} \frac{(u^i+u^{-i})(1-u^i)}{|1-u^i|} = \frac{\sin(3\pi/N)}{1-\cos(3\pi/N)} - \frac{\sin(\pi/N)}{1-\cos(\pi/N)} = b.
\end{equation}

This gives the expression for $R$ in the form

\begin{equation}
	R = \frac{N-1+K}{2a + Jb}.
	\label{eq.38}
\end{equation}

In Fig.\ \ref{Fig.10}(c) we plot the values of $R$ varying $J$ with $K=1.0$ (black curve). We also numerically calculate the values of $R$ for some values of $J$ (red dotted points) to show that numerical values replicate the analytical ones.

\section{\label{sec:level5}Swarmalators with the phase dynamics of the Stuart-Landau oscillator}
Having studied various emerging states of the swarmalators, we now want to find out what happens when a different kind of phase dynamics is considered in the swarmalator model in place of the vastly used Kuramoto-like dynamics. For each swarmalator moving in the two-dimensional plane, we associate a Stuart-Landau (SL) oscillator with it. The SL oscillator is an amplitude oscillator with two state variables and the dynamics is represented by

\begin{equation}
	\dot{\textbf{u}}_i = \begin{pmatrix} [1-(u_{i}^{2}+v_{i}^{2})]u_{i} - \omega_{i}v_{i}\\
		[1-(u_{i}^{2}+v_{i}^{2})]v_{i} + \omega_{i}u_{i} \end{pmatrix}\\ = F(\textbf{u}_i),
	\label{eq.23}
\end{equation}
where $\textbf{u}_i = (u_i,v_i)$ is the two-dimensional state variable and $\omega_i$ is the intrinsic frequency of the $i$-th swarmalator for $i = 1, 2, \ldots,N$. We consider symmetry preserving diffusive type coupling $K(\textbf{u}_i,\textbf{u}_j) = (u_j-u_i,v_j-v_i)^T$ for the interaction among the SL oscillators. As before, the spatial coupling is taken to be all-to-all, whereas we study only the local attractive coupling for the phase interaction between the swarmalators. Here we adopt the same coupling scheme as described in subsection~\ref{subsec:level2}.  The swarmalator model with Stuart-Landau oscillator is governed by the pair of equations
\begin{equation}
	\label{eq.24}
	\dot{\textbf{x}}_{i} = \frac{1}{N-1} \sum_{\substack{j = 1\\j \neq i}}^{N}\left[ \frac{\textbf{x}_{j}-\textbf{x}_{i}}{|\textbf{x}_{j}-\textbf{x}_{i}|} (1+J\cos(\theta_{j}-\theta_i))  -    \frac{\textbf{x}_{j}-\textbf{x}_{i}}{{|\textbf{x}_{j}-\textbf{x}_{i}|}^2} \right],
\end{equation}

\begin{equation}
	\label{eq.25}
	\dot{\textbf{u}}_i = \begin{cases}
		F(\textbf{u}_i) + \dfrac{\epsilon_a}{N_i} \sum_{\substack{j = 1\\j \neq i}}^{N} \text{A}_{ij} \frac{K(\textbf{u}_i,\textbf{u}_j)}{|\textbf{x}_{j}-\textbf{x}_{i}|} &\mbox{if } N_i \ne 0 \\ F(\textbf{u}_i) &\mbox{if } N_i = 0 \end{cases}
\end{equation}

where $J, \epsilon_a, N_i,$ and $\text{A}_{ij}$ are same as described in Section~\ref{sec:level2}. In Eq.\ \eqref{eq.24}, by the phase $\theta_i$ of the $i$-th SL oscillator, we mean the principal value of argument of the complex number $Z_i = u_i + l v_i$ ($l = \sqrt{-1}$), where $u_i,v_i$ are the state variables of the corresponding oscillator. Evidently, $\theta_i \in [-\pi, \pi)$, which fulfills the requirement of the phase variable of the swarmalators. We choose identical oscillators with $\omega_i = \omega = 3.0$. The initial conditions for the state variables are chosen randomly from $[-1,1]$.

\begin{figure*}[hpt]
	\centerline{
		\includegraphics[scale = 1.1]{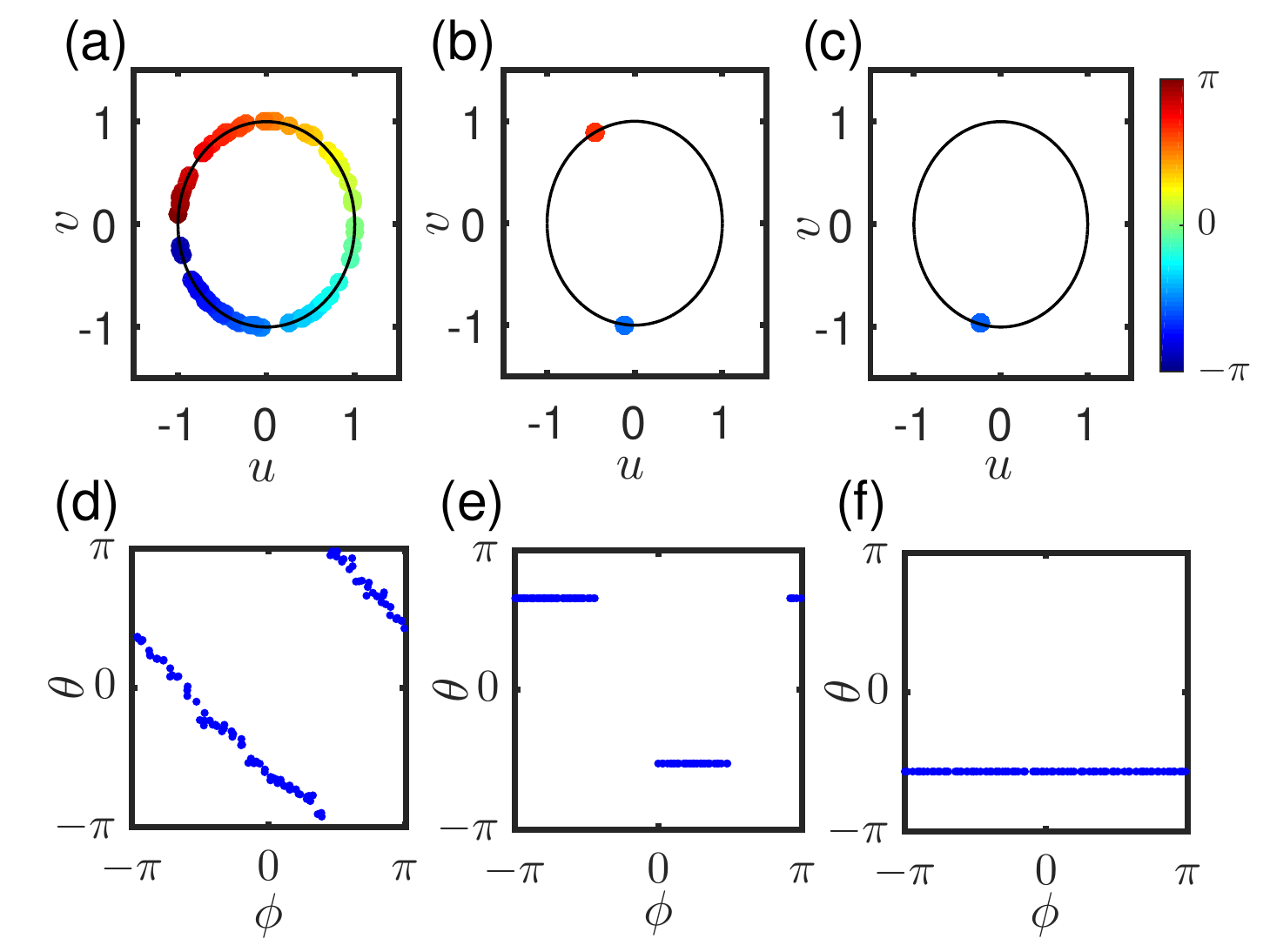}}					
	\caption{{\bf Swarmalators with the phase dynamics of SL oscillator}: Position in the $(u,v)$ plane of $N=100$ swarmalators at $t=300$ time unit in (a) static phase wave state for $r = 10^{-5}$, (b) static cluster state for $r = 0.3$, and (c) static sync state for $r = 1.0$, where $J = 0.8$. Color bar indicates the phase of the SL oscillators. The $\phi$-$\theta$ plots for the same values of $r$ and $J$ at the same time instant are shown in (d), (e), and (f), respectively. $\epsilon_a = 0.5$.}
	\label{Fig.9}
\end{figure*}

\par It is seen that all the states (namely, STPW, SCS, and SS), which have been found with Kuramoto oscillator with local attractive phase coupling, take place for SL oscillator as well. Attractive phase coupling invokes coherence among swarmalators' phases, and swarmalators in nearby phases attract themselves spatially with more strength, caused by the influence of phase similarity on spatial attraction. Due to the choice of non-zero intrinsic frequency, swarmalators' phases continue to evolve even when synchrony is achieved. Their positions in the $(x,y)$ plane become static as before since the difference of phase between any two swarmalators remains constant over time after transient period, and as a result, change of phase does not affect the spatial position. The Stuart-Landau oscillators execute limit cycle oscillation on a unit circle in the $(u,v)$ plane. In Figs.\ \ref{Fig.9}(a)-(c), positions of the oscillators on the limit cycle are shown for different emergent states, where they are colored according to their phase $\theta_i$. The relations between spatial angle $\phi$ and phase $\theta$ are shown in panel \ref{Fig.9}(d)-(f). For a very small choice of attractive vision radius $r$, phase coupling is effectively absent in the system and static phase wave (STPW) takes place. In the STPW state, the swarmalators' phases are distributed over $[-\pi,\pi)$, where their spatial angle $\phi$ and phase $\theta$ are perfectly correlated, which can be seen from the linear relation between $\theta$ and $\phi$ in Fig.\ \ref{Fig.9}(d). In the static cluster synchrony (SCS) state, the swarmalators form clusters both in $(x,y)$ (for position) and $(u,v)$ (for phase) plane. We show the formation of one such SCS state where the number of clusters is two (in Figs.\ \ref{Fig.9}(b) and \ref{Fig.9}(e)). The two horizontal bands in Fig.\ \ref{Fig.9}(e) clearly indicate that for the two clusters there are two different phases, whereas, their spatial angles are distributed over $[-\pi,\pi)$. The number of such clusters in the SCS state depends on the initial choices of spatial position and phase distribution of the swarmalators. The two clusters merge and form a single group when $r$ is increased and static sync (SS) state is achieved. It can be seen in Figs.\ \ref{Fig.9}(c) and \ref{Fig.9}(f) that all the oscillator' phases are same and as a result, the phase influenced spatial attraction among them increases to form a single cluster in space. Here, by using the SL oscillator for the phase dynamics of swarmalators, we validate the results in local attractive phase coupling scenario which we have found for swarmalators with the Kuramoto phase dynamics.

\section{Discussion and Conclusion} \label{sec:level6}

\par Swarmalators have been studied from the perspective of multi-agent systems for their position aggregation and phase dynamics. The bidirectional influence of agents' spatial position and internal phase on each other gives rise to fascinating collective patterns. Some of these stationary and non-stationary patterns are significantly found while studying the behaviors of real-world multi-agent systems. It is worth mentioning that in Ref.~\cite{lee2021collective} the collective behaviors of swarmalators have been studied with a finite-cutoff interaction distance, which is similar to the vision radius in our model. In this finite-cutoff interaction, the distance controlled spatial coupling among the swarmalators while keeping the phase coupling globally attractive or repulsive. Local spatial coupling among the swarmalators resulted in the formation of groups with spatially nearby ones. They found the emergence of multiple static sync discs (for globally attractive phase coupling) or multiple static async discs (for globally repulsive phase coupling).

\par In contrast, we have taken our spatial coupling to be all-to-all and phase coupling to be competitive and dependent on the vision radius. We have encountered cluster states in our model even when the spatial coupling is global. Swarmalators with same phase attract themselves and stay nearby in space. As a result, the cluster formation is influenced by the effect of phase similarity on spatial attraction rather than the direct local spatial coupling in Ref.~\cite{lee2021collective}. In another work \cite{hong2021coupling}, the coupling strength between the swarmalators was chosen randomly from a two-peaks distribution where the peaks correspond to positive and negative coupling strengths. The phase transition was studied while varying the probability of attractive phase coupling. The randomness in the phase coupling strength allowed the swarmalators to get repulsive or attractive coupling, irrespective of their spatial distance.  But, in our model, the coupling strategy is deterministic in the sense that swarmalators' phases are coupled attractively if they lie inside a specific vision range, and otherwise, the coupling is repulsive. While both these works represent competitiveness in the phase interaction, the underlying strategies and their emergent states are different.

\par In this article, we have modeled swarmalators in such a way that their phases are attractively coupled only when they lie inside each other's attractive vision range, otherwise their phases are coupled repulsively. It has been assured that the swarmalators avoid collisions in finite time and they maintain a minimal distance among them in space which is uniform in time. Our model comprises four parameters $r, J, \epsilon_a$, and $\epsilon_r$, which control the long-term behavior of the swarmalator system. Considering attractive vision radius $r$ to be very large and very small, phase coupling has been seen to become globally attractive and globally repulsive, respectively. A new static state (static cluster synchrony) where swarmalators form disjoint clusters is found, when $\epsilon_r$ is taken to be zero. We have analytically found a sufficient condition, which, when satisfied, leads to the static sync state. In the presence of attractive-repulsive phase coupling in the system, competitive phase interaction gives rise to several asymptotic states. Varying $r$, we have studied the transition of these states for different cases of $J, \epsilon_a$, and $\epsilon_r$ in Fig.\ \ref{Fig.8}. The two-clustered static $\pi$ state is a novel state for swarmalators in the two dimensional plane. This state is different from the static $\pi$ state found in Ref.~\cite{o2021collective} where swarmalators were positioned on a 1D ring. In the mixed phase wave states, the effect of competitive phase interaction is prominent where attractive phase coupling minimizes the phase difference between swarmalators at short distances and repulsive phase coupling increases the phase difference between the swarmalators who are outside each other's vision range. Four order parameters $R, S, T$, and $V$ unfold the qualitative behavior of the emergent dynamical states. To study swarmalators with a different oscillator, we have used the phase dynamics of the Stuart-Landau oscillator in our system and studied their behavior in the presence of local attractive phase coupling.

\par 
Our model has introduced a particular coupling strategy to incorporate competitive phase interaction. It remains to be seen what happens when this competitive interaction takes place following a different rule. In fact, a different set of values for the plausible choices of $\alpha$, $\beta$, and $\gamma$ make a significant variation in the observed dynamics of swarmalators, leading an important direction for future exploration. A step to study swarmalators with the Stuart-Landau oscillator in place of the conventional Kuramoto phase dynamics is considered in our work. Another scope of the study is to consider some other phase dynamics while modeling the swarmalator systems, resulting in different long-term behaviors. A possible step towards this direction is considering chaotic and hyperchaotic intrinsic dynamics \cite{sayeed2020behavioral,rossler1976equation,nag2020hidden,lorenz1963deterministic} with the positive Lyapunov exponent(s). This interesting inclusion may reveal further systematic insights; however, drawing a conclusion with chaotic dynamics requires more detailed rigorous analysis. The repulsive coupling may lead to an unbounded solution blowing out the dynamics away from the invariant manifold. It would be fascinating to investigate this future generalization expecting a broad spectrum of various novel dynamical states.

\par Moreover, we have considered the effect of phase similarity only on spatial attraction while studying the cumulative impact of attractive-repulsive interaction. It would be interesting if the effect of phase similarity on spatial repulsion is also considered under such competitive interactions.  Although we scrutinize the whole investigation with only swarmalators moving in two dimensions, the study on $3$D provides the same equations in (\ref{eq.3}-\ref{eq.4}), except the exponent of the repulsion will now be $\beta=3$ for a physically meaningful and analytically tractable model. We expect comparable results just like in $2$D. The static sync and async states will become spheres. We present a brief section devoted to understanding the influence of three-dimensional spatial movement in Appendix A \eqref{sec:level9}. A detailed analysis with three-dimensional spatial dynamics will be a theoretical avenue to explore. We believe that the results reported in this work with time-varying phase interaction will enrich the exiting study of swarmalators and emphasize competitive interaction in the swarmalator systems.

\color{black}
\section*{Acknowledgements} 

\par  We thank and gratefully acknowledge anonymous reviewers for their helpful comments and insightful suggestions that helped in considerably improving the manuscript. S.N.C acknowledges the financial support by the Council of Scientific \& Industrial Research (CSIR) under Project No.\ 09/093(0194)/2020-EMR-I. M.P. acknowledges funding from the Slovenian
Research Agency (Grant nos P1-0403, J1-2457, and J1-9112).

\section*{Data availability statement}

The data that support the findings of this study are openly available in the GitHub repository \cite{web_5}.

\section*{\label{sec:level9}Appendix A: $3$D model of swarmalators}

\begin{figure}[hpt]
	\centerline{
		\includegraphics[scale = 0.35]{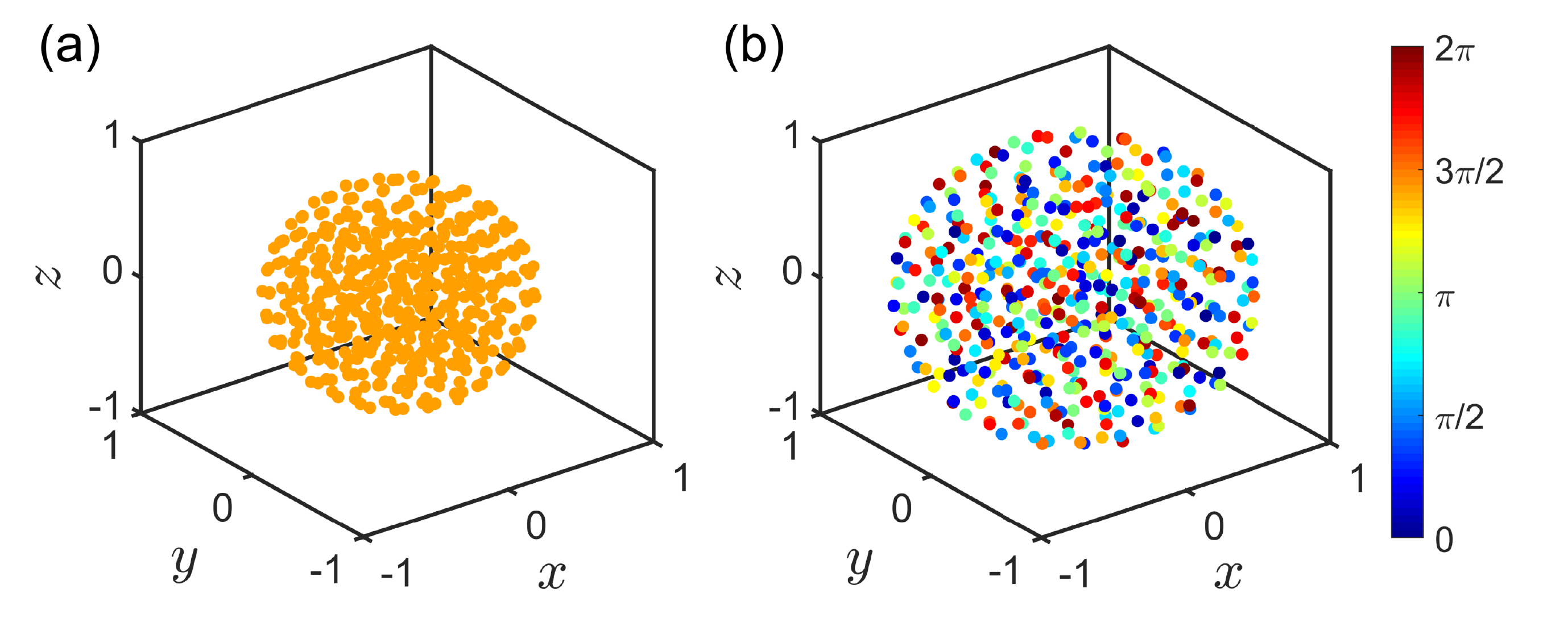}}
	\caption{{\bf Swarmalators in $3$D}: Snapshots of the swarmalators in three-dimensional space at $t = 200$ time unit. (a) Static sync state for $\epsilon_a = 0.5$, $\epsilon_r = -0.5$, $r = 10.0$, and $J = 0.8$. The static sync state takes spherical structure in $3$D where all the swarmalators lie inside this sphere and they are completely phase synchronized. (b) Static async state for $\epsilon_a = 0.5$, $\epsilon_r = -1.0$, $r = 10^{-5}$, and $J = 0.8$. In this state also the swarmalators form a spherical structure analogous to their $2$D disc shape where phases remain desynchronized. Here, $N=500$.}
	\label{Fig.12}
\end{figure}

One of the motivations of studying the dynamics of swarmalators is to mimic the behavior of real-world entities where external and internal dynamics are influenced by each other. Although the real-world systems are generally three-dimensional, for simplicity, we have introduced our model with two-dimensional spatial dynamics. After investigating the long-range behavior of the swarmalators under different interaction functions and coupling schemes, it is now suitable to find out if the results with $2$D spatial dynamics are generalized when the spatial dynamics are extended to $3$D. Now the spatial position of the $i$-th swarmalator is denoted by the vector $\textbf{x}_{i} = (x_i,y_i,z_i) \in \mathbb{R}^3$. In $3$D the swarmalator model is as follows

\begin{equation}
	\label{eq.41}
	\dot{\textbf{x}}_{i} = \frac{1}{N-1} \sum_{\substack{j = 1\\j \neq i}}^{N}\left[ \frac{\textbf{x}_{ij}}{|\textbf{x}_{ij}|} (1+J\cos(\theta_{ij}))  -    \frac{\textbf{x}_{ij}}{{|\textbf{x}_{ij}|}^{3}} \right],
\end{equation}

\begin{equation}
	\label{eq.42}
	\dot{\theta}_{i} = \frac{\epsilon_a}{N_{i}}\sum_{\substack{j = 1\\j \neq i}}^{N} \text{A}_{ij}\frac{\sin(\theta_{ij})}{|\textbf{x}_{ij}|^2} +  \frac{\epsilon_r}{N-1-N_{i}}\sum_{\substack{j = 1\\j \neq i}}^{N} \text{B}_{ij}\frac{\sin(\theta_{ij})}{|\textbf{x}_{ij}|^ 2},
\end{equation}
where $|\textbf{x}_{ij}|$ denotes the Euclidean distance between the $i$-th and the $j$-th swarmalators in three-dimensional space. These equations are similar to the governing equations of $2$D swarmalator model defined by Eqs.\ \ref{eq.3} and \ref{eq.4}. Although, here we have chosen  $\alpha = 1$, $\beta = 3$, and $\gamma=2$ which differ significantly from our previous choice of $\alpha = 1$, $\beta = 2$, and $\gamma=1$ for the earlier results of our work.


The states in the earlier simulations with the two-dimensional spatial dynamics now form their three-dimensional structures. In Fig.\ \ref{Fig.12}, we plot the snapshot of the swarmalators in the static sync and static async state. We find complete phase synchrony among swarmalators' phases in the static sync state where the two-dimensional disc structure is now converted into a three-dimensional sphere (see Fig.\ \ref{Fig.12}(a)). The same phenomenon takes in the static async state in Fig.\ \ref{Fig.12}(b) where swarmalators' phases remain desynchronized. The radii of these spheres depend on the choices of $\alpha$, $\beta$ and $\gamma$. All the other states found in the $2$D model can be found in the $3$D model too with appropriate choices of parameter values. Moreover, a different set of values for ($\alpha$, $\beta$, $\gamma$) does not change the qualitative behavior of the emerging asymptotic states significantly, as observed here.


\bibliographystyle{apsrev4-1} 
\providecommand{\noopsort}[1]{}\providecommand{\singleletter}[1]{#1}%

\end{document}